\documentclass[conference]{IEEEtran}
\IEEEoverridecommandlockouts

\usepackage{paralist}     
\usepackage{amsmath}       
\usepackage{amssymb}       
\usepackage{graphicx}     
\usepackage{textcomp}      
\usepackage{xcolor}        
\usepackage{url}          
\usepackage{enumitem}     
\usepackage{multirow}     
\usepackage{booktabs}      
\usepackage{listings}     
\usepackage{caption}       
\usepackage{subcaption}

\usepackage{tabularx}      
\usepackage{algorithm}     
\usepackage{algpseudocode} 
\usepackage[symbol]{footmisc} 
\usepackage{fancyhdr}    
\usepackage{comment}       

\usepackage{cite}         
\usepackage{textcase}      
\usepackage{yyy-math}

\usepackage{hyperref}

\hypersetup{
    pdfborder = {0 0 0}
}

\definecolor{dkgreen}{rgb}{0,0.6,0}
\definecolor{gray}{rgb}{0.5,0.5,0.5}
\definecolor{mauve}{rgb}{0.58,0,0.82}

\makeatother

\newcommand{\system}{{\sc Couler }}
\newcommand{\company}{{\sc Ant Group }}

\def\BibTeX{{\rm B\kern-.05em{\sc i\kern-.025em b}\kern-.08em
    T\kern-.1667em\lower.7ex\hbox{E}\kern-.125emX}}

\begin{document}

\title{Couler: Unified Machine Learning Workflow Optimization in Cloud\\
\thanks{Mingjie Tang, Yuan Tang, and Qian Jiang performed most of this work while at Ant Group.}
}

\author{
	 Xiaoda Wang\IEEEauthorrefmark{4}, Yuan Tang\IEEEauthorrefmark{2}, Tengda Guo\IEEEauthorrefmark{4}, Bo Sang\IEEEauthorrefmark{1}\\
	 Jingji Wu\IEEEauthorrefmark{1}, 
  Jian Sha\IEEEauthorrefmark{1},
  Ke Zhang\IEEEauthorrefmark{1}, Jiang Qian\IEEEauthorrefmark{6}, \IEEEauthorblockN{Mingjie Tang\IEEEauthorrefmark{4}
	}
    \IEEEauthorblockA{\fontsize{10}{10}\selectfont \IEEEauthorrefmark{1}\textit{Ant Group}~~~\IEEEauthorrefmark{2}\textit{Red Hat, Inc.}~~~\IEEEauthorrefmark{6}\textit{Snap, Inc}
    ~~~\IEEEauthorrefmark{4}\textit{Sichuan University}  \\
    \{wangxiaoda, guotengda\}@stu.scu.edu.cn, \{tangrock, terrytangyuan\}@gmail.com, \\\{b.sang, jingji.wjw, shajian, yingzi.zk\}@antgroup.com}
}

\maketitle

\begin{abstract}
Machine Learning (ML) has become ubiquitous, fueling data-driven applications across various organizations. Contrary to the traditional perception of ML in research, ML workflows can be complex, resource-intensive, and time-consuming.
Expanding an ML workflow to encompass a wider range of data infrastructure and data types may lead to larger workloads and increased deployment costs. 
Currently, numerous workflow engines are available (with over ten being widely recognized). This variety poses a challenge for end-users in terms of mastering different engine APIs. While efforts have primarily focused on optimizing ML Operations (MLOps) for a specific workflow engine, current methods largely overlook workflow optimization across different engines. 

In this work, we design and implement \system, a system designed for unified ML workflow optimization in the cloud. 
Our main insight lies in the ability to generate an ML workflow using natural language~(NL) descriptions. 
We integrate Large Language Models (LLMs) into workflow generation, and provide a unified programming interface for various workflow engines. This approach alleviates the need to understand various workflow engines' APIs. Moreover, \system enhances workflow computation efficiency by introducing automated caching at multiple stages, enabling large workflow auto-parallelization and automatic hyperparameters tuning. These enhancements minimize redundant computational costs and improve fault tolerance during deep learning workflow training.
\system is extensively deployed in real-world production scenarios at \company, handling approximately 22k workflows daily, and has successfully improved the CPU/Memory utilization by more than 15\% and the workflow completion rate by around 17\%.

\end{abstract}

\begin{IEEEkeywords}
Machine Learning Workflow, LLM, Cloud
\end{IEEEkeywords}

\section{Introduction}

A workflow, commonly known as a data pipeline, entails a sequence of steps that process raw data from various sources, directing it to a destination for both storage and analysis. Similarly, an ML workflow streamlines the comprehensive MLOps workflow, spanning data acquisition, exploratory data analysis (EDA), data augmentation, model creation, and deployment. Post-deployment, this ML workflow facilitates reproducibility, tracking, and monitoring. Such workflows enhance the efficiency and management of the entire model lifecycle, leading to accelerated usability and streamlined deployment~\cite{tfx, keystoneml}. To automate and oversee these workflows, ML orchestration tools are deployed, offering an intuitive and collaborative interface.

\begin{figure}[ht]
\vspace{-0.5em}
\centering
\includegraphics[width=1\columnwidth]{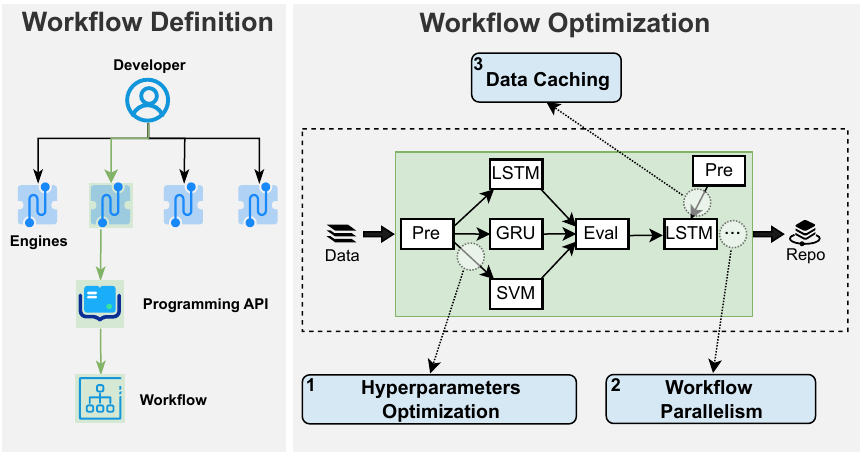}
\caption{An example of a financial company's journey in leveraging machine learning to predict market trends.}
\vspace{-0.5em}
\label{fig:workflow_example_intro}
\end{figure}

\noindent \textbf{Example.} Refer to the example in Figure~\ref{fig:workflow_example_intro}, a financial company aims to predict market trends using ML models. Initially, developers are tasked with selecting a workflow engine from a variety of available options, such as Argo, Airflow, Dolphin Scheduler, MetaFlow or Kubeflow Pipeline etc. Then, end users need to dedicate time to mastering the programming API of specific workflow engines. Upon defining the workflow, the first step entails data preprocessing. Subsequently, three models are evaluated, and the most promising model, LSTM, is selected for further analysis. The preprocessed data is reloaded for subsequent analysis, culminating in the generation of a predictive report through a complex process.
To implement this, the following challenges must be addressed:
\begin{itemize}
\item \textit{How can a workflow description be automatically translated to an ML workflow?} For execution, developers must code the workflow to be compatible with different workflow engines. However, the guidelines for different workflow engines can vary significantly, posing a challenge to become proficient in all of them.

\item \textit{How can the built workflow be effectively optimized?} Given a well-defined workflow, optimization is crucial. Developers need to find the optimal hyperparameters for training the ML models, and manage workflow parallelism manually. In the absence of caching, both the data loader and intermediate results become critical points, potentially slowing down the process.
\end{itemize}

\noindent \textbf{Goals and challenges.}
Given the ML workflow description and available resources, our objective is to autonomously construct a workflow that reduces dependence on expert knowledge. Simultaneously, we aim to enhance overall efficiency by minimizing end-to-end workflow execution costs. We strive to streamline the ML workflow creation process and ensure optimal utilization of available resources, making the entire system more user-friendly and efficient.

Effectively orchestrating workflows is crucial for companies heavily invested in machine learning. Consequently, a developer needs to understand the programming API of the selected workflow engine and learn to automate and optimize the entire workflow manually.
Numerous widely used workflow engines exist, such as Argo Workflows~\cite{Argo}, Tekton Pipelines~\cite{Tekton}, and Apache Airflow~\cite{Airflow}. The necessity to master multiple workflow engines presents a significant challenge for developers due to the unique programming interface of each engine.
With the advent of LLMs, significant strides has been made in the realms of natural language to SQL conversion~\cite{sun2023sql, rajkumar2022evaluating, gao2023text}, code generation from natural language descriptions~\cite{nijkamp2022codegen, xu2022systematic} and database performance tuning~\cite{lao2023gptuner}.
This advancement facilitates the efficient conversion of natural language descriptions into programming coding across different workflow engines, thereby simplifying the workflow definition process.
However, several challenges remain:

Given the myriad of available workflow engines, attempting a direct translation from NL to various workflow engine codes proves to be intricate and inefficient. Factors such as the continual evolution of workflow engine APIs and the distinct design philosophy behind each engine contribute to this complexity. Additionally, LLMs may not always stay updated with the latest changes in these APIs, posing a challenge to ensure accurate NL to code translation consistently. This scenario accentuates the need for a unified coding interface catering to different workflow engines. Such an interface simplifies the process of defining and managing workflows without delving into the intricacies of each engine, thereby enhancing the efficiency of LLMs in translating NL descriptions into executable code.

After establishing a workflow, optimizing its computational aspects is crucial. One challenge is to effectively cache intermediate results dynamically, maximizing resource use and minimizing runtime. Storing crucial intermediary outputs allows workflows to gracefully handle runtime errors without the need to restart from scratch. Moreover, splitting large workflows into smaller, more manageable segments is not straightforward. It demands careful strategizing to strike a balance between performance and resource use. In ML workflows, hyperparameter optimization of the models introduces another layer of complexity.
Identifying the optimal hyperparameter values is a complex process, and leveraging the capabilities of LLMs to automate this process, while promising, remains a significant challenge.

\noindent \textbf{Contributions.} To address these challenges, 
the contributions of this work are outlined below:

\begin{itemize}[leftmargin=*,itemindent=-0em]

\item \textbf{Simplicity and Extensibility}: We provide a unified programming interface for workflow definition, ensuring independence from the workflow engine and compatibility with various workflow engines such as Argo Workflows, Airflow, and Tekton. We demonstrate how \system supports ML model selection and AutoML pipelines.

\item \textbf{Automation}: We integrate LLMs in unified programming code generation. By leveraging LLMs, we facilitated the generation of unified programming code using NL descriptions. Additionally, we automate hyperparameters tuning through the integration of Dataset Card and Model Card, enhancing the effectiveness of the autoML process.

\item \textbf{Efficiency}: We introduce the Intermediate Representative (IR) to depict the workflow Directed Acyclic Graph (DAG), optimizing extensive workflow computations by dividing a large workflow into smaller ones for auto-parallelism optimization. We also implement dynamic caching of artifacts, which are the outputs of jobs in the workflow, to minimize redundant computations and ensure fault tolerance.

\item \textbf{Open Source Community}: We constructed the platform to assist data scientists in defining and managing workflows, enabling system deployment in real production environments on a large scale. The released open-source version has garnered adoption from multiple companies and end-users\footnote{\url{https://couler-proj.github.io/couler/}}. For instance, over 3000 end users are utilizing \system within \company, and more than 20 companies have adopted \system as their default workflow engine interface.

\end{itemize}

\section{System framework}
\label{sec:system_overview}

\begin{figure*}
\centering
\includegraphics[width=\textwidth]{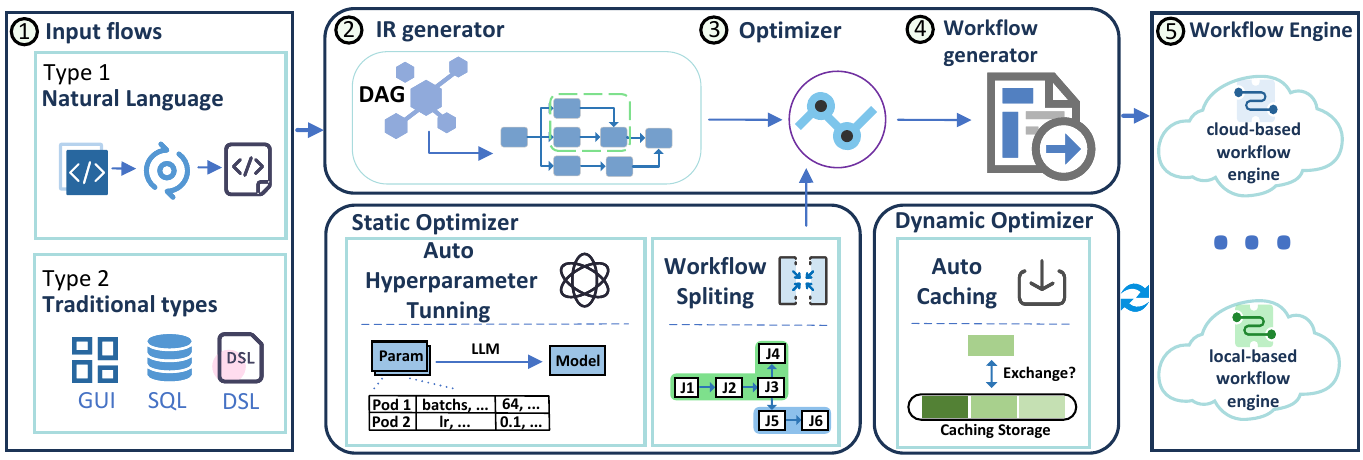}
\caption{Overview of \system~Architecture.}
\vspace{-1.5em}
\label{fig:system_framework}
\end{figure*}

Figure~\ref{fig:system_framework} illustrates the \system architecture, highlighting various components and multiple aggregation layers that facilitate scaling across clusters.
Initially, we provide two interfaces for defining workflows: one through Natural Language and the other through GUI, SQL, and programming languages such as Python and GoLang. 
Once a workflow is defined, it's converted into an Intermediate Representation (IR) format. Subsequently, optimization measures, specifically the auto hyperparameter tuning optimizer and workflow auto-parallelism, are employed to refine the workflow. Upon completion, \system generates the final workflow which is then submitted to the designated workflow engine. Concurrently, an automated caching mechanism operates in real-time, dynamically updating the cache as the workflow progresses.

\vspace{-0.8em}
\subsection{Workflow Description}

We offer two primary methods for users to construct workflows. The first leverages Natural Language (NL) descriptions, wherein we employ LLMs, such as ChatGPT-3.5 and ChatGPT-4, to generate code compliant with a standardized workflow interface definition~\hyperref[nltocoding]{(\S \uppercase\expandafter{\romannumeral3})}. Simultaneously, users can alternatively create workflows using a Graphical User Interface (GUI)\hyperref[sec:implementation]{(\S \uppercase\expandafter{\romannumeral5})}, SQL tools like SQLFlow\hyperref[sec:implementation]{(\S \uppercase\expandafter{\romannumeral5})}, or directly through programming languages such as Golang or Python.

\vspace{-0.35em}
\subsection{Workflow DAG Generator}
\label{unifiedprogramminginterface}
We propose a unified programming interface to define workflows in a DAG way. This interface is designed to allow users to delineate workflows without specific knowledge of the underlying workflow engine. And it offers fundamental functions such as executing scripts, containers, or jobs, stipulating conditions, and managing multiple instances of a job, among others. 
For example, the code~\ref{lst:dag_example_full} shows how to build a workflow implicitly. By this way, users need to own a clear big picture for the workflow, and under how the running logic among steps in their real application. The definition of DAG workflow via explicit way helps data engineer to debug a failed workflow more easily, and build a complicated workflow with hundred nodes. 
More detailed information about the interface is given in Appendix A.

\begin{lstlisting}[caption={An Example Workflow DAG in \system}, label={lst:dag_example_full}, frame=bt]
def job(name):
    couler.run_container(
        image="whalesay:latest",
        command=["cowsay"],
        args=[name], step_name=name)

def diamond():
    couler.dag(
    [[lambda: job(name="A")],  
        [lambda: job(name="A"), # A -> B
            lambda: job(name="B")],  
        [lambda: job(name="A"), # A -> C
            lambda: job(name="C")],  
        [lambda: job(name="B"), # B -> D
            lambda: job(name="D")],  
        [lambda: job(name="C"), # C -> D])
            lambda: job(name="D")]
    ])
diamond() /*Execute the diamond function.*/
submitter = ArgoSubmitter()
/*Submit and run the workflow over Argo.*/
couler.run(submitter=submitter)
\end{lstlisting}
\vspace{-1em}

\vspace{-0.35em}
\subsection{Workflow Intermediate Representation}

A workflow processes a stream of input data to train a model, subsequently generating a new model for machine learning applications. Typically, a workflow is represented in a DAG format. Consequently, we represent a workflow in an intermediate representation (IR) format, unbound to any specific backend workflow engine or platform. Utilizing IR allows us to optimize the workflow independently of platform-related properties, enabling \system to assimilate workflows from the unified programming interface.

\vspace{-0.35em}
\subsection{Auto Tuning Optimizer and Workflow Optimizer}

We utilize LLMs to generate recommended hyperparameter configurations for machine learning models, by analyzing dataset characteristics from Dataset Card and model information from Model Card~\hyperref[autohyper]{(\S \uppercase\expandafter{\romannumeral4}.C)}. This approach automates the fine-tuning of hyperparameters in machine learning workflows, enabling LLMs to generate configurations that enhance model performance.
Based on the workflow's IR, the \system server employs a rule-based approach to formulate the optimization plan before initiating a workflow. The considerations for this plan include optimizing large workflows, resource request optimization, and the reuse of intermediate results. All optimizations adhere to a predefined interface, incorporating their specific implementations. Further details regarding these optimizations are provided in Section~\hyperref[bigworkflow]{(\S \uppercase\expandafter{\romannumeral4}.B)}.

\vspace{-0.35em}
\subsection{Automatic Caching Optimizer}

In \system, artifacts are integrated as valuable products of workflow development, including datasets, parameters, diagrams, etc. Various physical storage options are available and can be registered to accommodate different types of artifacts. We offer an Automatic Caching Mechanism based on the artifact to dynamically update the cache during workflow execution~\hyperref[autocachingm]{(\S \uppercase\expandafter{\romannumeral4}.A)}. For each currently executing pod, a comprehensive analysis is conducted across three dimensions: past usage, future usage, and the cost-effectiveness of caching. This analysis yields a cache value score, used to re-evaluate the existing cache content. This re-evaluation helps determine whether updates need to be made to the cache.

\vspace{-0.35em}
\subsection{Workflow Generator and Workflow Engines}

\system aims to enable workflows to operate across various platforms, with a particular focus on cloud-native processing. To accelerate execution, we aim to support workflow generation tailored to specific platforms. As a result, the final phase of \system optimization involves generating workflows to execute on distinct workflow engines. The workflow generator converts the intermediate representation of a DAG to an executable format. Then, a workflow engine like Argo can execute this format (e.g., YAML format for Argo workflow). 
This YAML is then sent to the Argo operator within a Kubernetes cluster, demonstrating how the abstraction of IR allows for flexibility in supporting various workflow engines. In Kubernetes, the workflow engine operates as a workflow operator. Initially, this operator allocates the associated Kubernetes resources (i.e., Pods) according to the resource definition for a step in a workflow, and then monitors the status of steps, updating the workflow status as needed. The execution topology of the workflow is dictated by the workflow's DAG, with the workflow operator scheduling the relevant steps in the cluster based on the status of steps and the DAG.

\section{NL to Unified Programming Interface}
\label{nltocoding}

In this section, we explore the application of LLMs for converting Natural Language (NL) to Unified Programming Interface as shown in Section\hyperref[unifiedprogramminginterface]{(\S \uppercase\expandafter{\romannumeral3}.B)}. Traditional methods involve defining workflows using various techniques and submitting them to a cluster. Lately, LLMs have demonstrated remarkable performance across a wide array of inference tasks. However, upon direct application of LLMs for unified programming code generation, certain challenges arise: Firstly, the overall workflow complexity hampers the performance of LLMs in complete workflow conversion. Secondly, LLMs possess limited knowledge regarding \system's unified programming interface.

To address these challenges, we introduce a method that leverages LLMs to automatically translate natural language into unified programming code via the crafting of task-specific prompts. This approach enables users to articulate their desired workflows in natural language, which are then automatically translated into executable unified programming code. As a result, our method simplifies the \system workflow creation process and improves usability for individuals with limited programming experience, as illustrated in Figure~\ref{fig:automl_code_method}. We also introduce this procedure through a running example in Section~\hyperref[nltowexper]{(\S \uppercase\expandafter{\romannumeral5}.D)}. The transition from NL descriptions to \system code encompasses four pivotal steps:

\begin{figure}[ht]
\centering
\includegraphics[width=0.95\columnwidth]{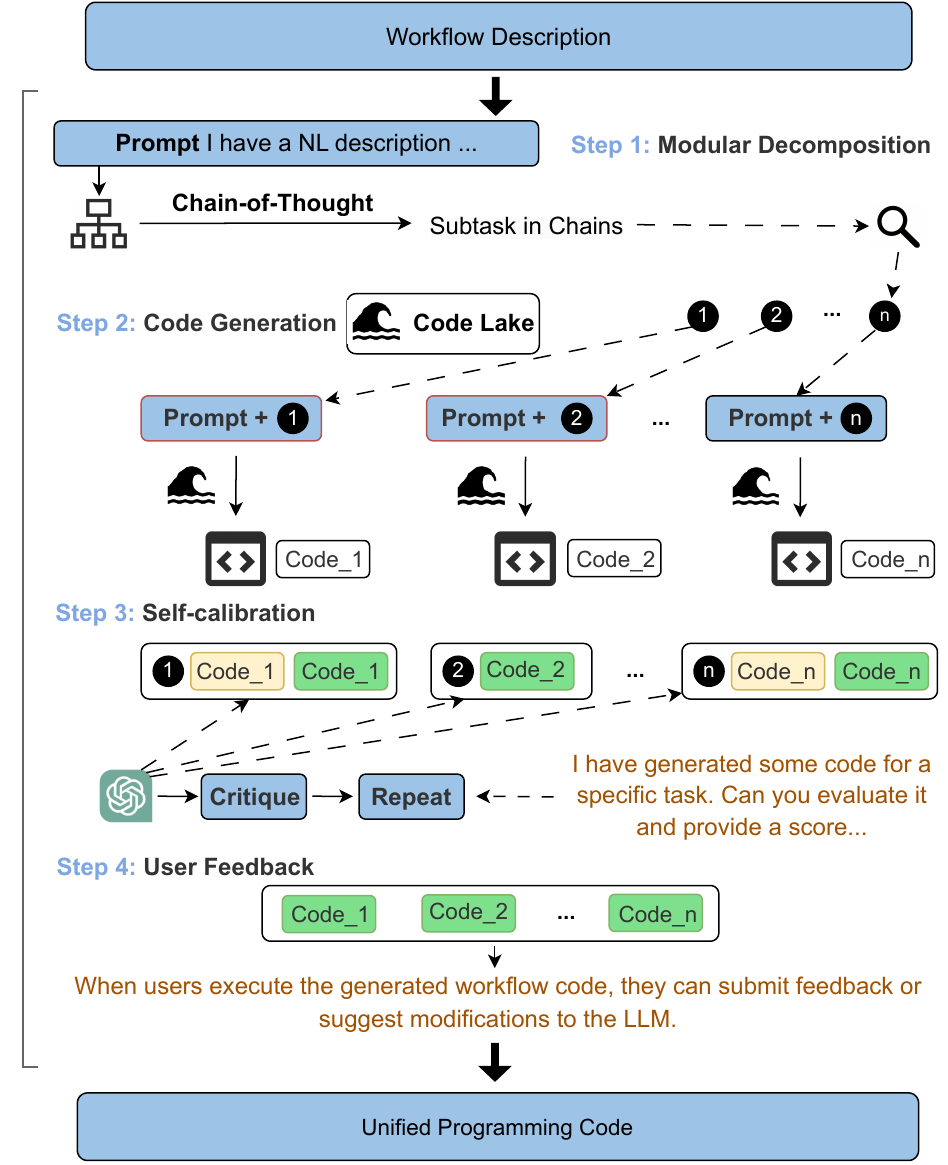}
\caption{NL to Unified Programming Interface}
\vspace{-1.0em}
\label{fig:automl_code_method}
\end{figure}

\noindent \textbf{Step 1: Modular Decomposition}: Initially, we employ a chain of thought strategy~\cite{wei2022chain} to decompose natural language descriptions into smaller, more concise task modules, such as data loading, data processing, model generation, and evaluation metrics. Each module should encapsulate a singular, coherent task to ensure the precision and correctness of the generated \system code. A series of predefined task types can be established to identify and extract pertinent tasks based on the input of natural language descriptions automatically. They provide a structured approach to ensure the precision and correctness of the code generated.

\noindent \textbf{Step 2: Code Generation}: For each independent subtask, we utilize LLMs to generate code. Considering that LLMs have limited knowledge about \system, we construct a Code Lake containing code for various functions. We search for relevant code from the Code Lake for each subtask and provide it to LLMs for reference. This significantly improves the ability for unified programming code generation.
\begin{algorithm}
    \caption{NL to Unified Programming Interface}  
    \begin{algorithmic}[1]
    \newcommand{\Input}{\item[\textbf{Input:}]} 
    \newcommand{\Output}{\item[\textbf{Output:}]}    
    \Input Description $\mathcal{D}$, LLM $\mathcal{L}$, {Baseline\_Score} $S_b$
    \Output Executable \system Code $\mathcal{C}$
    
    \State \textbf{Modular Decomposition:} \textbf{chain of thought} to decompose NL description $\mathcal{D}$ into smaller chains $d_i$
    
    \For{each subtask $d_i$ in chains}
        \State \textbf{Generate Subtask Code:}
        \State \quad Search for relevant code as reference
        \State \quad Use $\mathcal{L}$ to generate code $c_i$ for the subtask $d_i$
        \State \textbf{Self-calibration:} 
        \State \quad Compute score $s_i$ for $c_i$ leveraging $\mathcal{L}$
        
        \While{$s_i$ $<$ $S_b$}{}
            \State Re-generate subtask code and update $s_i$
        \EndWhile
    \EndFor
    \State \textbf{User Feedback:} review and validate the generated unified programming code
    \end{algorithmic}
    \label{methodforselfllm}
\end{algorithm}

\noindent \textbf{Step 3: Self-calibration}: After generating the code for each subtask, we integrate a self-calibration strategy~\cite{tian2023just} to optimize the generated code. This strategy evaluates the generated code by having LLMs critique it, as shown in Algorithm~\ref{methodforselfllm}. 
In line 8, there may be complex scenarios in which achieving the desired score is impractical for various reasons. Users can adjust $Baseline\_Score$ in instances where it is set too ambitiously, rendering it unattainable.
Initially, we define a baseline score \(S_b\) as the standard evaluation score. We use LLMs to evaluate the generated code \(c_i\) for a score \(s_i\) between 0 and 1, and if \(s_i < S_b\), we will provide feedback of LLMs and repeat the code generation. After this self-calibration, we will have improved code for each subtask.

\noindent \textbf{Step 4: User Feedback}: Finally, users can review and validate the generated workflow code. If the generated code fails to meet the users' requirements, they have the opportunity to provide feedback and suggestions in textual format. The system will leverage this feedback to optimize the code and enhance the precision of code generation.

\section{Workflow Optimization}
\label{sec:optimizer}

In this section, we present three optimizations implemented at \company to enhance workflow efficiency. Firstly, we introduce an artifact auto-caching mechanism to eliminate redundant computations. Secondly, for workflows comprising thousands of nodes, we propose a heuristic approach to partition large workflows into smaller units, thereby maximizing workflow parallelism. Lastly, we introduce an automatic hyperparameters tuning method based on LLMs to automate the training pipeline of ML workflows. 

\subsection{Automatic Artifact Caching Mechanisms}
\label{autocachingm}

\begin{table}[h]
\centering
\vspace{-1em}
\caption{Set of common notations used in our description.}
\begin{tabularx}{0.45\textwidth}{lX}
   \toprule
   \multicolumn{1}{c}{\textbf{Notation}} & \textbf{Definition} \\
   \midrule
   $G$  &  DAG of workflow $(G= \left \langle J, E, C \right \rangle )$ \\
      &  \qquad  Jobs J, Edges E, Configurations C \\
   $A$  & Adjacency matrix of a directed graph G \\
   $J_s, J_p$ & Serial and Parallel Job Sets \\
   $u$  &  Artifact $u$\\   
   $\mathcal{L}$  & The reconstruction cost of artifacts \\
   $\mathcal{F}$  & The utility value of artifacts\\
   $\mathcal{V}$  & The cache cost of artifacts\\
   $\mathcal{I}$ & The cache assessment metrics of artifacts\\
   $N_c$  & List of cached artifacts: $\{u_1, ..., u_i\}$ \\
    $t, s$ & Computation Time and space usage of Jobs\\
   \bottomrule
\end{tabularx}
\end{table}

\textbf{Motivation of Caching.} 
Machine learning model development is a highly iterative process, often involving repeated steps with variations. This iterative nature can lead to significant duplicated work, especially concerning data import and transformation. By caching intermediate results, such as preprocessed data or feature representations, data scientists can avoid redundant computations across iterations, thus accelerating the development process. This increased iteration speed translates into higher productivity, allowing problems to be solved faster and empowering data scientists. However, caching introduces additional overhead e.g., storage costs. In this work, we introduce the way to strike the right balance between storage overhead and computational cost savings in \company.

\subsubsection{Problem Statement and Evaluation Metrics}

Caching all intermediate data (called artifacts in this work) is an instinctive approach, but it comes with challenges. Firstly, not all data merits caching, especially if it is not slated for reuse in the foreseeable future. Secondly, the associated costs of caching can be prohibitive. For instance, at \company, we delegate intermediate artifact storage to distributed in-memory systems like Apache Alluxio\cite{li2018alluxio}. Given the finite memory capacities of such systems, making judicious decisions about which artifacts to cache is crucial. This necessitates an automatic selection mechanism that factors in the caching expense cost when determining which data to store.

Thus, we prioritize workflow execution time and memory consumption as the pivotal performance metrics and targets for optimization. Specifically, we define the workflow execution time, represented as \(\mathcal{T}\), as the duration required for completing the Critical Path. This Critical Path is characterized as the elongated sequence of interdependent tasks spanning from the inception to the culmination of the workflow as in~\cite{spark-nsdi}. On the other hand, the metric for memory expenditure, symbolized as \(\mathcal{S}\), is construed as the peak memory consumption observed across all concurrently operating nodes. Based on these definitions, the cost function can be articulated as follows:

\begin{equation}
    \label{argmin}
    \mathcal{T} = \max ( \sum_{p \in J_t} t_{p} )
\end{equation}
\begin{equation}
    \label{argmin2}
    \mathcal{S} = \max ( \sum_{p \in J_s} s_{p} )
\end{equation}

where $t_{p}$ and $s_{p}$ is the time and memory usage for Job $p$. We define the job groups with the longest running time and the largest resource consumption as $J_t$ and $J_s$, respectively.

\subsubsection{Principles of Automatic Caching}

In this study, we propose a metric called the \textit{caching importance factor} to ascertain the significance of caching a specific artifact (namely $u$). This factor serves as a guiding principle to dynamically determine which artifact warrants caching. We represent this by a function, $\mathcal{I}(u)$, which computes the \textit{caching importance factor} for artifact $u$. 
Our formulation of this metric is primarily influenced by three determinants: the cost of reconstructing the artifact, denoted as $\mathcal{L}$; the expected value of reusing the artifact, represented as $\mathcal{F}$; and the associated expense of caching, labeled $\mathcal{V}$. Details are presented below.  

\textbf{Artifact reconstruction cost:} refers to the expense incurred when re-creating or regenerating machine learning artifacts or intermediate results that were not cached or saved during the workflow. 
When these artifacts are not cached or saved, and they need to be reconstructed from raw data or recomputed, it can result in additional computational expenses, increased execution time, and potentially higher resource usage. Minimizing artifact reconstruction costs is one of the objectives of effective caching strategies in ML workflows.

In this research, given an artifact $u$, we focus on analyzing the subgraph containing nodes that serve as predecessors to artifact $u$, which we refer to as $G_p=\{J_1, ..., J_s\}$.
Note that, to simplify our discussion in this work, we only consider subgraphs with the following properties: (a) We select the subgraph $G_p$, formed by the preceding $n$ layers of jobs from node $u$, as it is the most representative. (b)If the artifact of a job within $G_p$ is cached, $G_p$ will be truncated at that point.
On this basis, we hope to minimize the related artifact reconstruction cost $\mathcal{L}(u)$. Within this subgraph $G_p$, the cost $\mathcal{L}(u)$ is determined by the computational resources utilized by jobs and the storage resources associated with the artifacts involved. Formally, $\mathcal{L}(u)$ is defined as follow way: 
\begin{equation}
\label{eq0}
\mathcal{L}(u) = \sum_{i=1}^{s} \sum_{j=1}^{s} A_{ij} \cdot (w_{i} + d_i \cdot d_j)
\end{equation}
where, $A$ denote the adjacency matrix, respectively. $w_{i}$ represents the resource consumption of job $i$. 
The degree $d_i$ indicates the level of significance for job $i$, and $s$ represents the number of nodes in $G_p$. By this way, we formulate the overall runtime complexity of $G_p$ , taking into account the varying importance of each node.

\textbf{Artifact reuse value:} 
refers to the benefits and advantages gained by reusing previously generated artifacts (e.g., preprocessed data, feature representations, or model checkpoints) in a machine learning workflow. The value comes from avoiding redundant computations and leveraging the work done in earlier stages of the workflow, ultimately leading to resource savings and more efficient model development. Maximizing the reuse value is another optimization target in this work. 

Given an artifact $u$, the artifact reuse value name as $\mathcal{F}(u)$ is influenced via the successor of workflow graph. This graph is referred as $G_s$ whose definition is the same as $G_p$. Within this subgraph $G_s=\{J_1, ..., J_t\}$, we hope to maximize the artifact reuse value $\mathcal{F}(u)$ as following way. 
\begin{equation}
\label{eq2}
\mathcal{F}(u) = \sum_{i=1}^{t}{\frac{r}{\kappa_{ui}} \cdot (\zeta_{ui}+1)}
\end{equation}
Where $\kappa_{ui}$ represents distance for node $u$ and node $i$ in the subgraph $G_s$, $r$ represents a boolean state indicating whether a reuse event occurs for artifact $u$ and $t$ represents the number of nodes in $G_s$. Then, $\zeta_{ui}$ is the weighted value for the dependency of job $i$ on $u$.
Given the adjacency matrix as $A$ and the degree of nodes as $d$. We use $diag$ to represent the diagonal matrix, and matrix $\zeta$ can be computed as follow:
\begin{equation}
\label{eq3}
\zeta = diag[d_1, ..., d_n] - A
\end{equation}

\textbf{Artifact caching cost:}  refers to the expenses associated with storing and managing cached artifacts or intermediate results in a machine learning workflow. In this work, we use the distributed in-memory storage to store the artifact, thus, we mainly consider $u$'s memory consumption (name as $\mathcal{V}(u)$).

Overall, given a artifact $u$, we formalize the \textit{caching importance factor} of $u$ as follow:
\begin{equation}
\label{eq5}
\mathcal{I}(u) = \alpha\cdot\log(1 + \mathcal{L}(u)) + \beta\cdot\mathcal{F}(u)^2 - e^{-\mathcal{V}(u)}
\end{equation}
where $\mathcal{\alpha}$ and $\mathcal{\beta}$ are weight parameters for the metrics, and their optimal values are selected through experimental studies in the production environment.
The $\mathcal{\alpha}$ and $\mathcal{\beta}$ are used to adjust the weights among the three factors: reconstruction cost, reuse value, and cache cost. As the impact of these factors on efficiency varies in different training scenarios, it is necessary to adjust them according to the actual situation.

\begin{algorithm}
\caption{Automatic Caching Mechanisms}
\label{alg:auto-caching-process}
\begin{algorithmic}[1]
\State \textbf{Input:} JobSet $\mathcal{N}$, Workflow $\mathcal{G}$, Artifact Cached List $N_c$, Used Caching Storage $C_u$, Total Caching Storage $C_t$
\State \textbf{Output:} Dynamic Caching Set $D_c$
  
\State \textbf{function} $\mathcal{L}$($u$) $\to$ Returns artifact reconstruction cost of $u$
\State \textbf{function} $\mathcal{F}$($u$) $\to$ Returns artifact reuse value of $u$
\State \textbf{function} $\mathcal{V}$($u$) $\to$ Returns artifact caching cost of $u$
\State \textbf{function} $\mathcal{I}$(${l}$, ${f}$, ${v}$) $\to$ Returns \textit{caching importance factor} of $u$
    
\State $C_u \gets \emptyset$, $N_c \gets \emptyset$
\State markUnVisited($\mathcal{G}$)
    
\ForAll{$u \in \mathcal{N}$}
    \If{not Visited($u$) and $C_u < C_t$}
        \State $u \rightarrow N_c$
    \ElsIf{not Visited($u$) and $C_u \geq C_t$}
        \State NodeSelection($u$, $\mathcal{G}$, $N_c$, $C_u$, $C_t$)
    \EndIf
\EndFor
    
\Function{NodeSelection}{$u$, $\mathcal{G}$, $N_c$, $C_u$, $C_t$}
    \ForAll{$u \in N_c$}
        \State $v_i \gets$ $\mathcal{V}$($u$) \Comment{$\mathcal{V}$($u$):}memory consumption
        \State $l_i \gets$ $\mathcal{L}$($u$) \Comment{using Eq.~(\ref{eq0})}
        \State $f_i \gets$ $\mathcal{F}$($u$) \Comment{using Eq.~(\ref{eq2})}
        \State $I_i \gets$ $\mathcal{I}$($l_i$, $f_i$, $v_i$) \Comment{using Eq.~(\ref{eq5})}
    \EndFor
     
    \State MarkVisited($u$)
    
    \While{$C_u > C_t$}
        \State $u_{min} \gets \arg\min_{u_i \in N_c} I_i$
        \If{$u_{min} \neq u$}
            \State $u \text{ in } N_c$, $u_{min} \text{ out } N_c$
        \Else
            \State $u_i \text{ out } N_c$
        \EndIf
        \State update $C_u$
    \EndWhile
\EndFunction

\end{algorithmic}
\end{algorithm}

The \textit{caching importance factor} plays a crucial role in deciding whether a new artifact should replace an existing one in the cache memory. This factor is instrumental in enabling \system to maximize execution time efficiency while working within the constraints of limited cache space. 
We will recompute the caching importance factor of all remaining items in the Caching Storage whenever an item is removed.
In this way, we hope to reduce the communication overhead in the workflow and the reconstruction cost when artifacts are reused, thereby decreasing the overall runtime $\mathcal{T}$. We introduce the Algorithm~\ref{alg:auto-caching-process} to determine which artifacts should be cached during the caching process based on the constraint. 

To make optimal cache exchange decisions, \system's dynamic caching module calculates the caching value of newly generated artifacts during the workflow execution process.  Algorithm~\ref{alg:auto-caching-process} provides an overview of how the dynamic caching strategy module makes cache decisions and optimizes execution time efficiency. The monitor attempts to place newly generated artifact into the cache (line 11). If there is insufficient cache space, we calculate a cache score based on the attributes of the new artifact (line 16-21). 
This score is then compared to the scores of artifacts already in the cache (line 24-30), determining whether to remove an existing cached artifact. This process is repeated until there is enough cache storage available or the score of the new artifact is lower than the compared score.

\vspace{-5mm}

\begin{figure}[h]
\centering
\includegraphics[width=0.9\columnwidth]{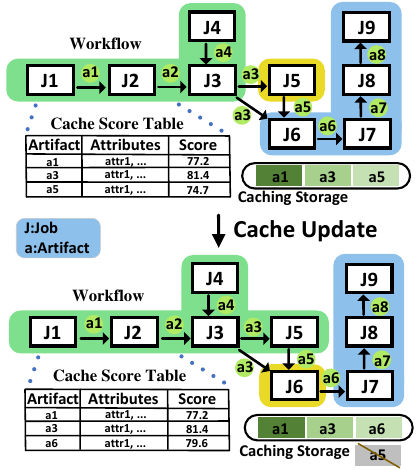}
\caption{Running Example of Automatic Caching}
\vspace{-0.5em}
\label{fig:example}
\end{figure}

\subsubsection{Running Example of Automatic Caching}

Figure~\ref{fig:example} presents a running example for the caching strategy in this work. In the workflow, the green sections represent the Jobs that have completed execution, the yellow sections indicate the Jobs currently in execution, and the blue sections denote the Jobs awaiting execution. Arrows represent the dependency relationships between Jobs. The Cache Score Table maintains records of the size, type, and other attributes of cached artifacts, as well as their cache scores. Note that caching storage refers to the cache space allocated for the workflow. 

Upon the completion of $J6$, \system calculates the cache score for artifact $a6$ based on the attribution of $a6$. Subsequently, \system attempts to store artifact $a6$ in the Caching Storage. If the remaining space is to be insufficient for $a6$, \system compares $a6$’s cache score with that of $a5$, which has the lowest score in the Cache Score Table. Due to $a6$’s score is higher than $a5$, $a5$ is replaced by $a6$. If the Caching Storage is still inadequate, the comparison continues with the artifact having the next lowest score, and this process is repeated until an artifact with a higher score than $a6$ or adequate storage capacity becomes available to cache $a6$.

\subsection{Big Workflow Auto Parallelism Optimization}
\label{bigworkflow}

In general, a workflow can be very big (i.e., more than one thousand nodes). At \company, we run into the case where the workflow involves more than four hundred nodes. This would bring two issues. At first,  each workflow is a Kubernetes CRD (Custom Resource Definition), the CRD is defined in YAML format and the size of CRD is limited to specific requirements. For example, the API server of Kubernetes would be overflowed by the large CRD (e.g., the size of YAML can not bigger than 2MB in practice). Secondly, the user cannot define the workflows properly to achieve maximum parallelism in a big DAG, therefore, the optimizer of \system needs to analyze the dependence of workflow and split the workflow into multiple ones. 

\begin{algorithm}
\caption{Big Workflow Auto Parallelism Mechanisms}
\label{alg:workflow-auto-parallel}
\begin{algorithmic}[1] 
\newcommand{\Input}{\item[\textbf{Input:}]} 
\newcommand{\Output}{\item[\textbf{Output:}]}
\Input Budget $\mathcal{C}$, Workflow $\mathcal{G}$
\Output Multiple split workflows $W_s$
\State Cand $\gets \emptyset$, $ W_s \gets \emptyset$
\State markUnVisited($\mathcal{G}$)
\ForAll{$n_i \in \mathcal{N}$}
    \If{not Visited($v_i$)}
        \State NodeSelection($n_i$, $\mathcal{G}$, $N_c$, $C_u$, $C_t$)
    \EndIf
\EndFor
\Function{SplitWorkflow}{$v_1$, $\mathcal{C}$, $\mathcal{G}, W_s, Cand$}
    \State $b_i \gets$ $BudgetOnUnVisitedVertex$($\mathcal{G}$)
    \If{($b_1 \leq \mathcal{C}$)}
        \State $W_s \gets W_s + \mathcal{G}$
        \State \textbf{return}  \text{$W_s$}
    \EndIf
    \State MarkVisited($v_i$)
    
    \State $\overline{C}$ $\gets$ Cand + $v_1$,
         $b_2 \gets$ BudgetOnGraph( $\overline{C}$) 
    \If{$b_2 \ge \mathcal{C}$}
        \State $W_s \gets W_s + $ Cand, 
        Cand $\gets v_1$ 
    \Else
        \State Cand $\gets$ $\overline{C}$
    \EndIf
    \ForAll{$v \in adj(v_1)$}
        \If{not Visited($v$)}
            \State SplitWorkflow($v$, $\mathcal{C}$, $\mathcal{G}, W_s$, Cand)
        \EndIf
    \EndFor
\EndFunction

\end{algorithmic}
\end{algorithm}

In this paper, we first define the budget of workflow. The budget is used to decide whether we need to split a big workflow into small ones. The budget (namely $\mathcal{C}$ ) could be the (a)  size of workflow CRD in YAML format: ($\alpha$), (b) the  number of steps in a workflow: $\beta$, (c) the  number of pods: ($\gamma$) in a workflow. Thus, $\mathcal{C} = \alpha + \beta + \gamma $. 
In this work, we mainly use the size of workflow $\alpha$ as the default budget value. For example, $\alpha$ exceeds 2 MB or $\beta$ exceeds 200.
Naturally, if a workflow is bigger than a predefined budget, it needs to be split into small ones. 

Given the required budget and a big workflow in DAG format, the optimization goal is a problem of finding optimal DAG sets to schedule workflow so we can win the maximum parallel. It is tempting to reach for classical results ~\cite{Topological} in the optimal graph topological order to identify an optimal schedule. The topological ordering of a directed graph could be used to split a big graph into smaller graphs for scheduling. In this work, we identify a workflow sets by depth-first search (DFS) over a DAG. Algorithm~\ref{alg:workflow-auto-parallel} goes through each vertex of the graph and puts this vertex into a workflow candidates greedily until each vertex is visited or the workflow meeting the budget requirement. Initially, we mark every vertex as unvisited in line 1 and recursively split the related DAG from the unvisited vertex one by one from lines 2 to 4. Function $SplitWorkflow$ is used to split the input DAG. At first, we check whether the current workflow meets the requirement, that is, the budget is smaller than the requirement from lines 7 to 9.  Next, we mark the current vertex $v_1$ as visited and check whether it is possible to add the vertex $v_1$ into the DAG candidate $\overline{C}$. if the vertex $v_1$ fails to join the current subgraph $\overline{C}$, we put the current subgraph $\overline{C}$ into the output set of DAGs (namely $W_s$). Finally, we go through the adjacent list of $v_1$ and continue to split the input DAG. Function $BudgetOnUnVisitedVertex$ in line ~7 and $BudgetOnGraph$ in line ~10 compute the related budget for the input graph for the un-visited vertex of input DAG or the whole DAG, respectively. Because we go through the input DAG via the depth first search order, the runtime cost of the proposed approach is the number of vertex (i.e., $O(|V|)$).

\vspace{-0.3em}
\subsection{Automatic Hyperparameters Tuning}
\label{autohyper}

We explore the use of LLMs for automatic hyperparameters tuning of machine learning models by analyzing dataset characteristics from Dataset Card~\cite{gebru2021datasheets} and model information from Model Card~\cite{mitchell2019model}. This approach automates the fine-tuning of hyperparameters in machine learning workflows, enabling LLMs to generate configurations that enhance model performance. We detail the implementation approach and demonstrate how this automated configuration process improves the efficiency and effectiveness of model training in Algorithm~\ref{alg:autohcadasd}.

\begin{algorithm}[ht]
    \caption{Automatic Hyperparameters Tuning}
    \label{alg:autohcadasd}
    \begin{algorithmic}[1]
    \newcommand{\Input}{\item[\textbf{Input:}]} 
    \newcommand{\Output}{\item[\textbf{Output:}]}    
    \Input Data Card $\mathcal{D}$, Model Card $\mathcal{M}$, Hyperparameters Set $\mathcal{H}$, LLM $\mathcal{L}$
    \Output Targeted Hyperparameters $h_t$
    
    \State \textbf{Data Card:} comprise of the dataset name, input dataset type, label space, and default evaluation metrics 
    \State \textbf{Model Card:} consist of the model name, model structure, model descriptions, and architecture hyperparameters
    
    \For{each hyperparameters $h_i$ in $\mathcal{H}$}
        \State \textbf{Predicted Training Log:} 
        \State /* Generate a training log $t_i$ for a given hyperparameter setting $h_i$ by leveraging LLM $\mathcal{L}$. */
    \EndFor
    \State \textbf{Targeted Hyperparameters:}  
    \State \quad $h_t$ $\gets$ best performance for $h_i$ in $\mathcal{H}$ based on $t_i$ 
    \end{algorithmic}
\end{algorithm}

To fully exploit the capabilities of LLMs and generate effective prompts, we tailor prompts to the Data Card, Model Card, and hyperparameters information. The Data Card \( \mathcal{D} \) comprehensively describes the dataset, including details such as data name, data type, label space, and evaluation metrics. The Model Card \( \mathcal{M} \) provides a thorough description of the model, encompassing the model name, structure, description, and architecture hyperparameters, while the hyperparameters cover various parameter value ranges.

Initially, we have a hyperparameters set \( \mathcal{H} \) containing a few optional hyperparameters. Without requiring training on actual hardware, we employ LLMs to automatically predict performance during the training process, subsequently returning a training log for each hyperparameters \( h_i \) in \( \mathcal{H} \)~\cite{zhang2023automl}. This log captures various parameters and information from the training process. By examining the training log, we can observe the effects of different hyperparameters during training. After several rounds of testing, we select the training hyperparameters that yield the best performance. We introduce this procedure through a running example in Appendix C. 

\section{Implementation}
\label{sec:implementation_1}

The Python SDK for \system is now open-source. Several top enterprises have integrated this SDK into their production environments. The whole \system service is crafted in Golang, encompassing all internal components. 
Regarding the expressiveness of \system's API compared to the complete APIs of the supported workflow engines, \system has achieved over 90\% coverage of the Argo API. Additionally, it supports approximately 40-50\% of the Airflow API. We are actively working to enhance our support for Airflow and other workflow engines.
Due to space constraints, we put extensive discussions on implementation of \system into the Appendix B. 

\section{Evaluation}
\label{sec:experiment}

\begin{figure*}[ht]
    \centering
    \begin{subfigure}{0.32\textwidth}
        \includegraphics[width=\linewidth]{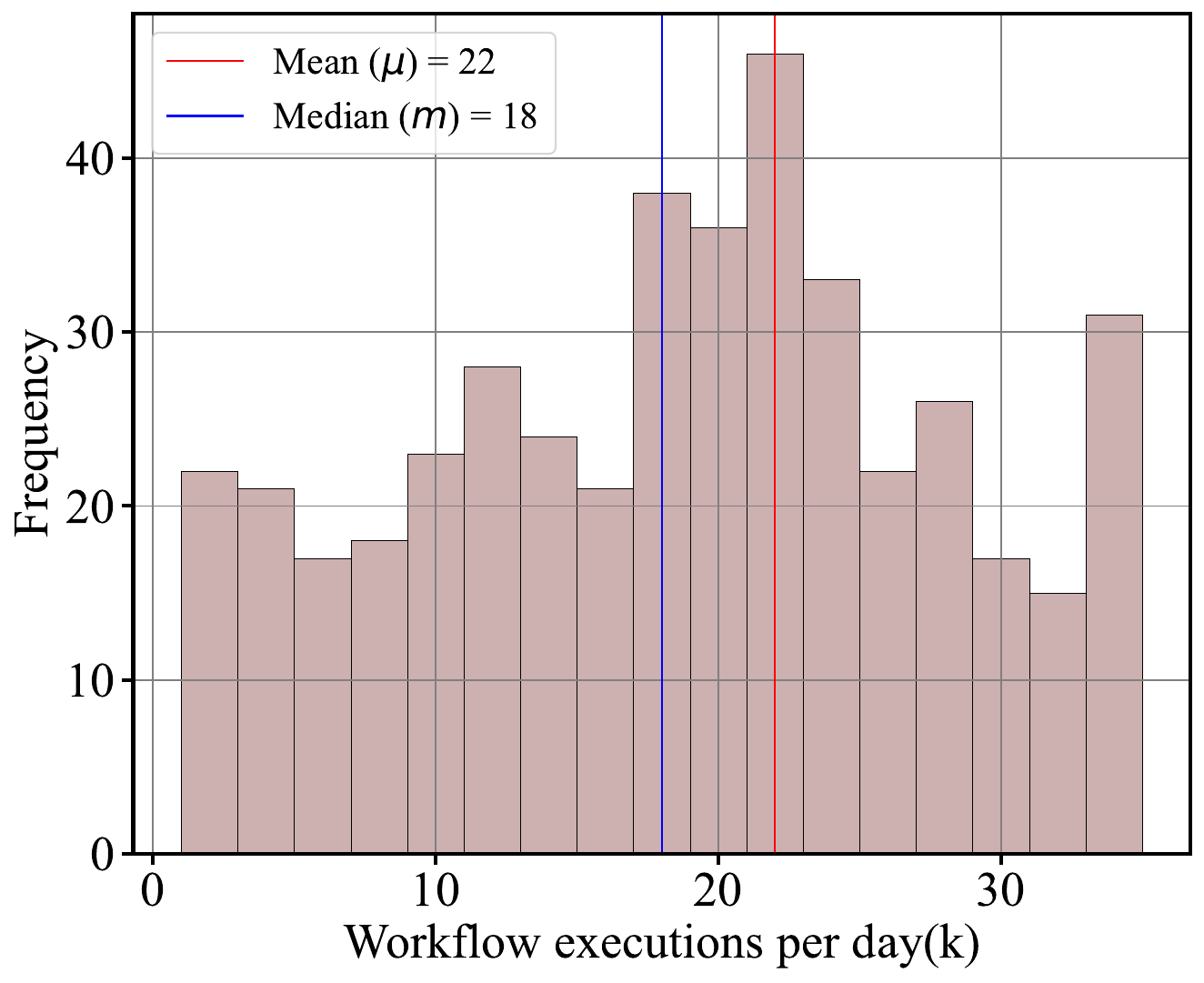}
        \caption{Distributin of workflow quantity}
        \label{fig:workflownumber}
    \end{subfigure}
    \hfill
    \begin{subfigure}{0.32\textwidth}
        \includegraphics[width=\linewidth]{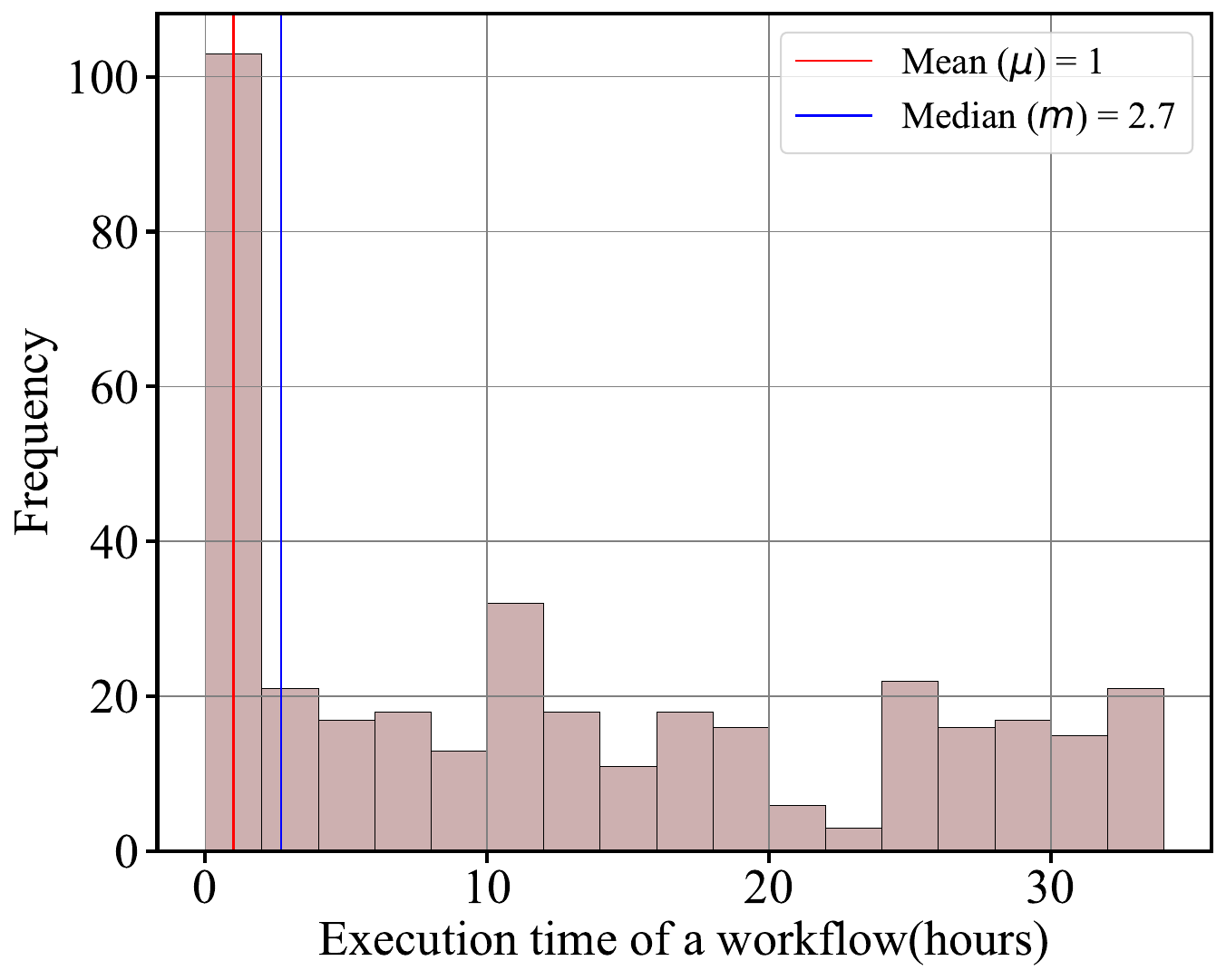}
        \caption{Distributin of workflow lifespan}
        \label{fig:workflowlifetime}
    \end{subfigure}
    \hfill
    \begin{subfigure}{0.32\textwidth}
        \includegraphics[width=\linewidth]{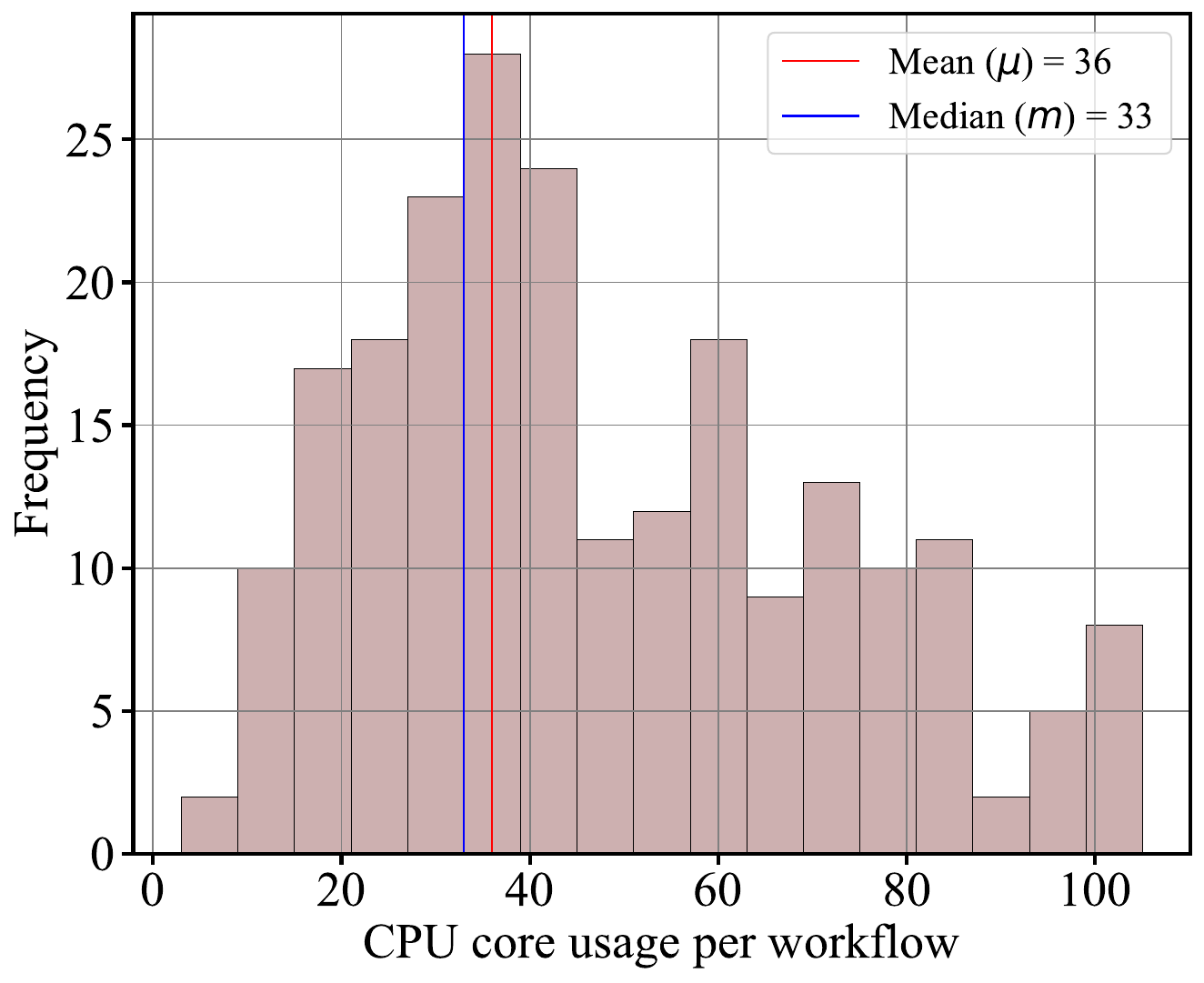}
        \caption{Distributin of CPU core usage}
        \label{fig:workflowcore}
    \end{subfigure}
    \caption{From July 2022 to July 2023, workflow activity analysis of \system in \company}
    \label{fig:workflowacticity}
     \vspace{-1.0em}
\end{figure*}

\begin{figure*}[ht]
    \centering
    \begin{subfigure}{0.32\textwidth}
        \includegraphics[width=\linewidth]{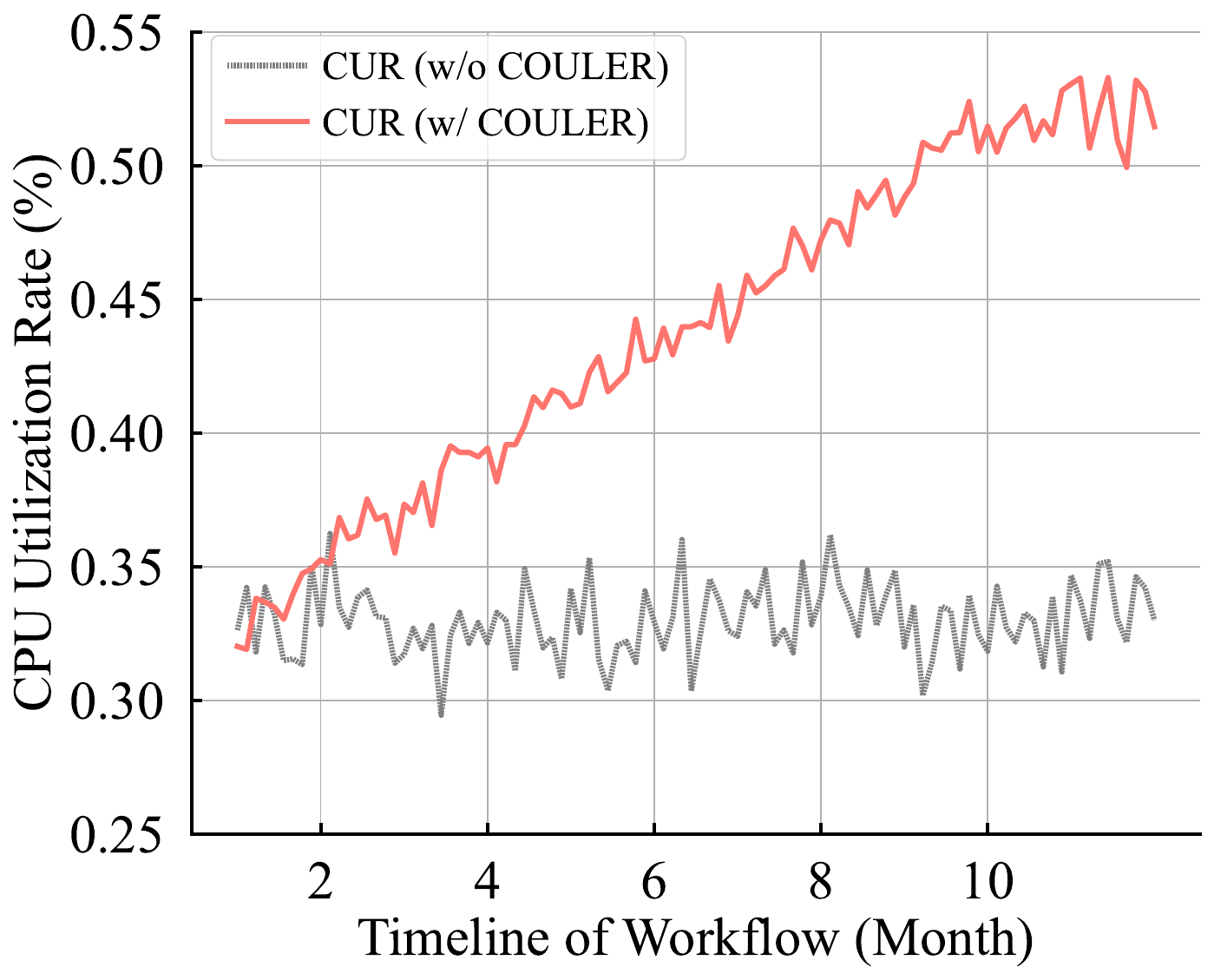}
        \caption{Yearlong evolution of CUR}
        \label{fig:coulercur}
    \end{subfigure}
    \hfill
    \begin{subfigure}{0.32\textwidth}
        \includegraphics[width=\linewidth]{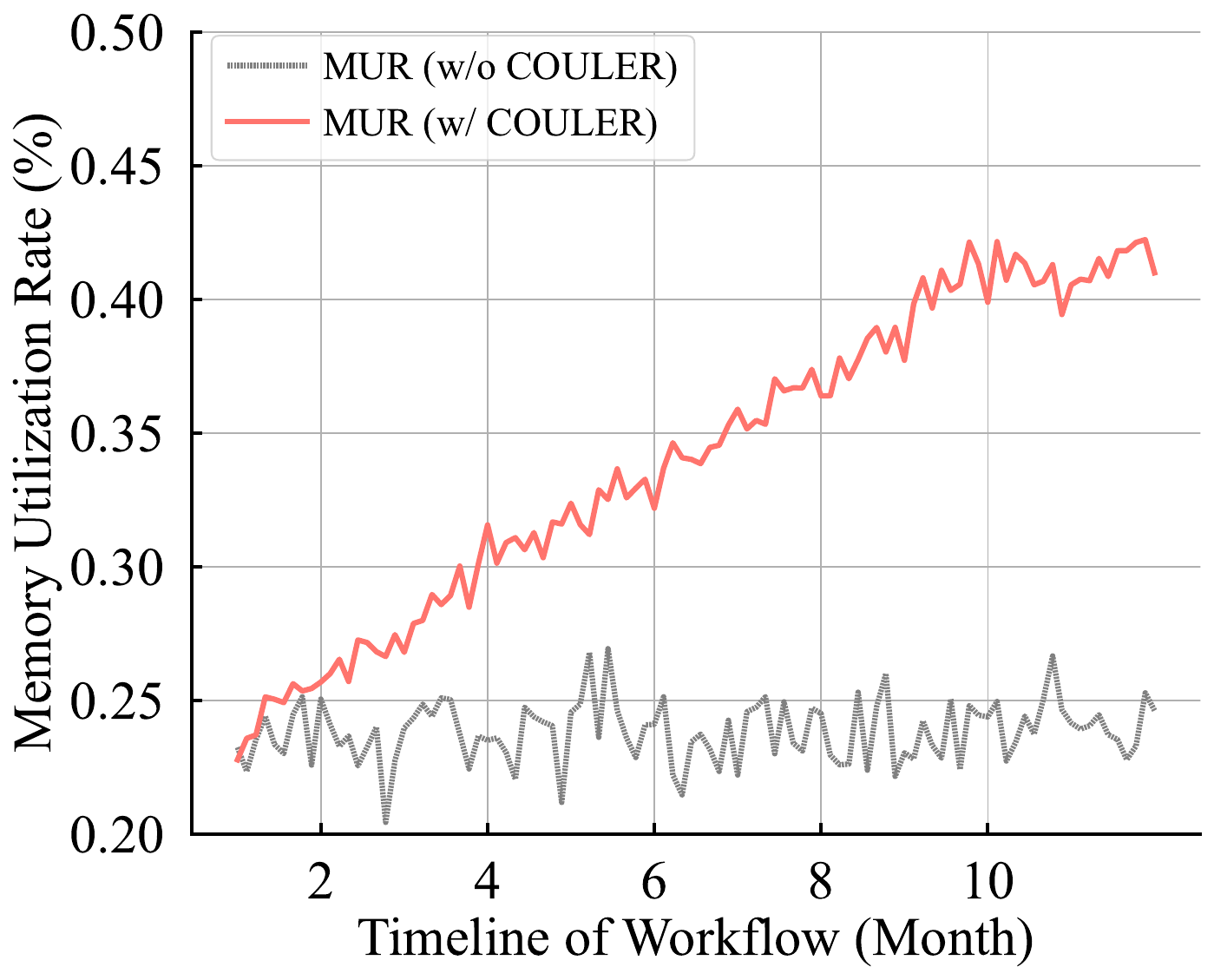}
        \caption{Yearlong evolution of MUR}
        \label{fig:coulermur}
    \end{subfigure}
    \hfill
    \begin{subfigure}{0.32\textwidth}
        \includegraphics[width=\linewidth]{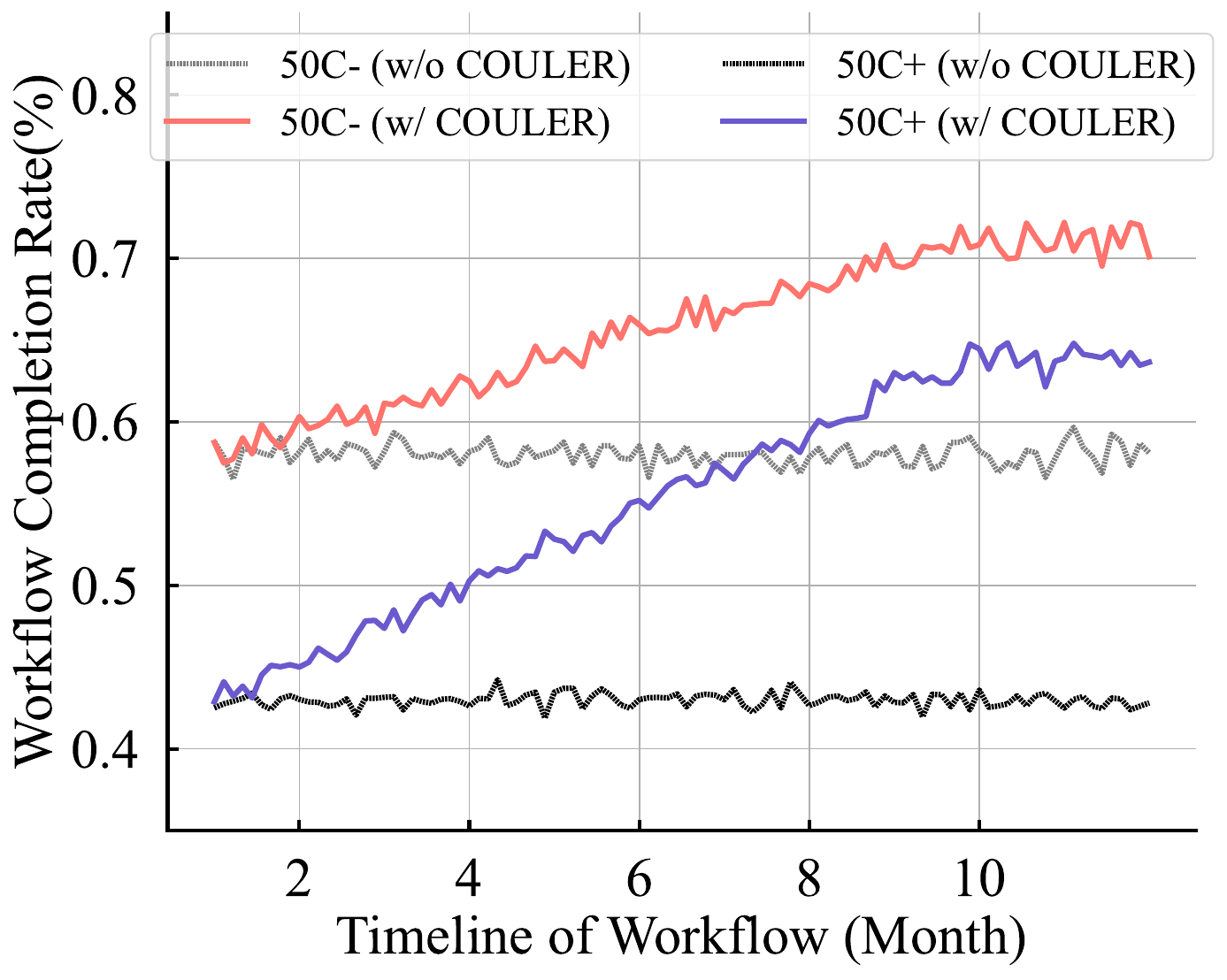}
        \caption{Yearlong evolution of WCR}
        \label{fig:coulerjcr}
    \end{subfigure}
    \caption{From July 2022 to July 2023, 90\% workflows in the cluster were transitioned to utilize \system in \company}
    
    \label{fig:coulerantgroup}
     \vspace{-1.5em}
\end{figure*}

%

Our evaluation results, which encompass diverse industrial models and data spanning several months, aim to address the following research questions:
\begin{itemize}[leftmargin=*,itemindent=-0em]
    \item \textbf{RQ1:} What is the usage frequency of \system in \company?
    \item \textbf{RQ2:} How effective is the automatic caching performance of \system?
    \item \textbf{RQ3:} How about the performance of NL to Unified Programming Code Generation?
    \item \textbf{RQ4:} How proficient is \system's capability of automatic hyperparameter configuration?
\end{itemize}

\subsection{Experiment Setup}

\noindent \textbf{Production Environment.}
In \company, various types of workflows operate concurrently in a shared cluster. The cluster provides substantial resources, with about 1,600,000 CPU cores, 4,500 GPU cores, 3.24 PB of memory, and 344 PB of disk space. This setup supports \company's diverse computational needs, enabling different workflows to run efficiently within the same shared resource environment. 
\system is utilized to support over 95\% of workflows in the production environment (e.g., 22k/day). The extensive scale of workflow operations provides accurate statistical estimates of the actual gain.

\noindent \textbf{Workload.}
We design a multi-modal workflow in an isolated production environment to minimize interference of the production environment and conduct a more comprehensive assessment of \system's capabilities. By selecting appropriate component containers for model training, we evaluated the system's caching efficiency as well as the performance of its AutoML features. This workflow comprises two distinct tasks: image classification and language model fine-tuning. We tested the performance of models such as ViT and nanoGPT. Additionally, the workflow incorporates system testing modules and model update modules, increasing the task complexity to better emulate real-world scenarios. The workflow includes 26 different training scenarios and comprises 52 working pods, utilizing over 1.4 million images and 20GB of text data as datasets. Operating in contexts with a significant number of parameters and data volume, it effectively showcases \system's unique features.

\vspace{-0.5em}
\subsection{Workflow Activity: RQ1}

\begin{figure*}[ht]
    \centering
    \begin{subfigure}{0.32\textwidth}
        \includegraphics[width=\linewidth]{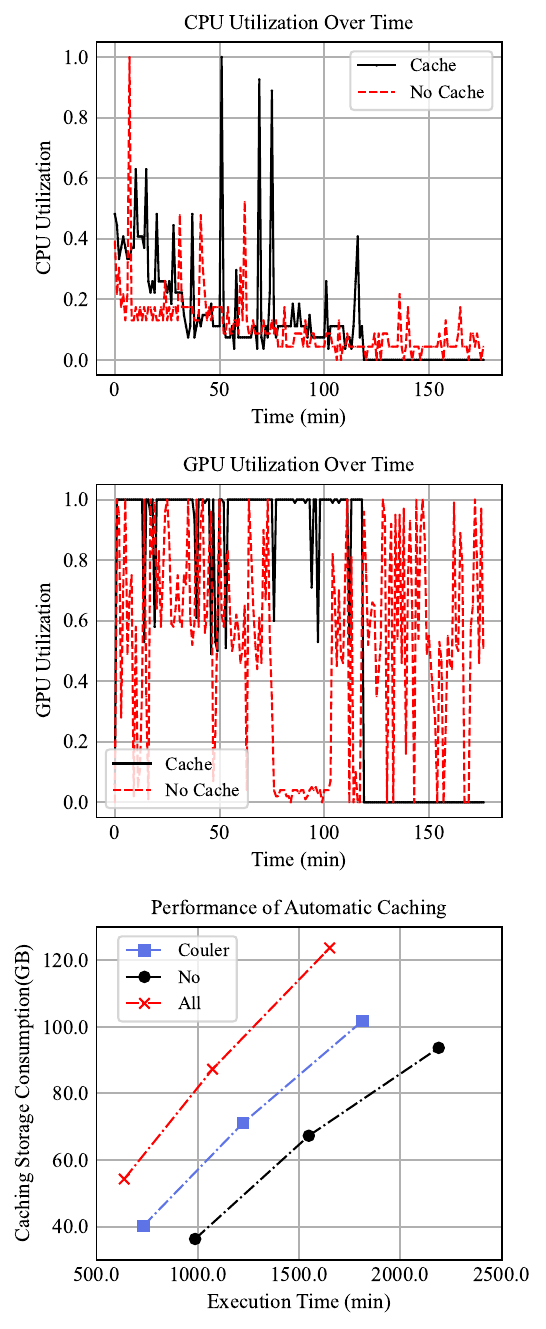}
        \caption{Multimodal Training}
        \label{fig:multimodal}
    \end{subfigure}
    \hfill
    \begin{subfigure}{0.32\textwidth}
        \includegraphics[width=\linewidth]{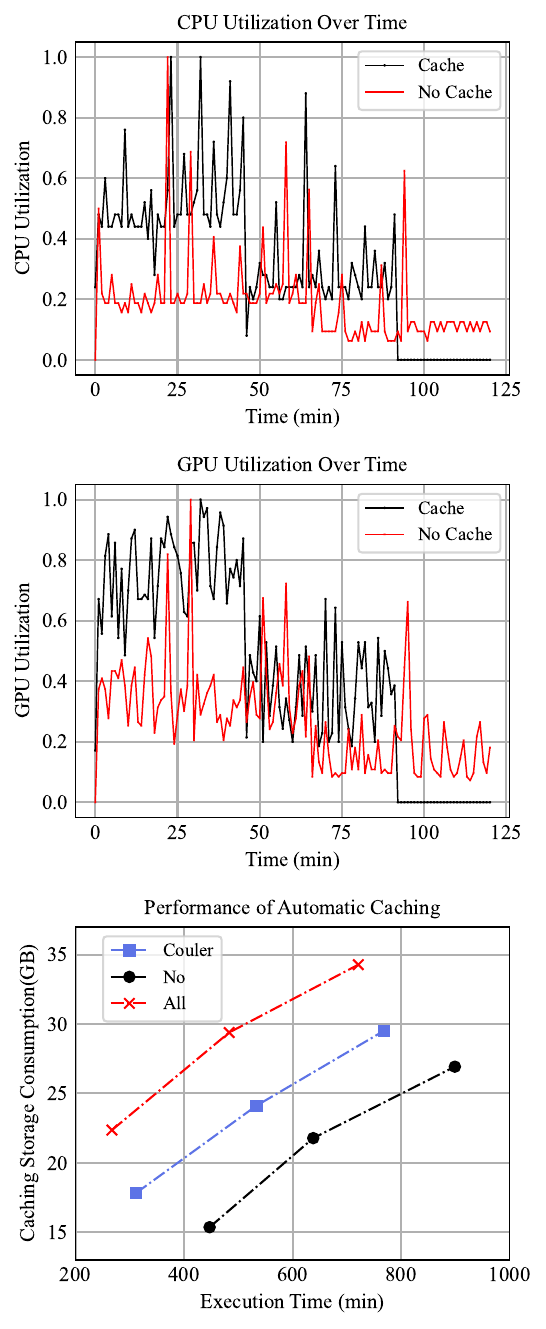}
        \caption{Image Segmentation}
        \label{fig:image}
    \end{subfigure}
    \hfill
    \begin{subfigure}{0.32\textwidth}
        \includegraphics[width=\linewidth]{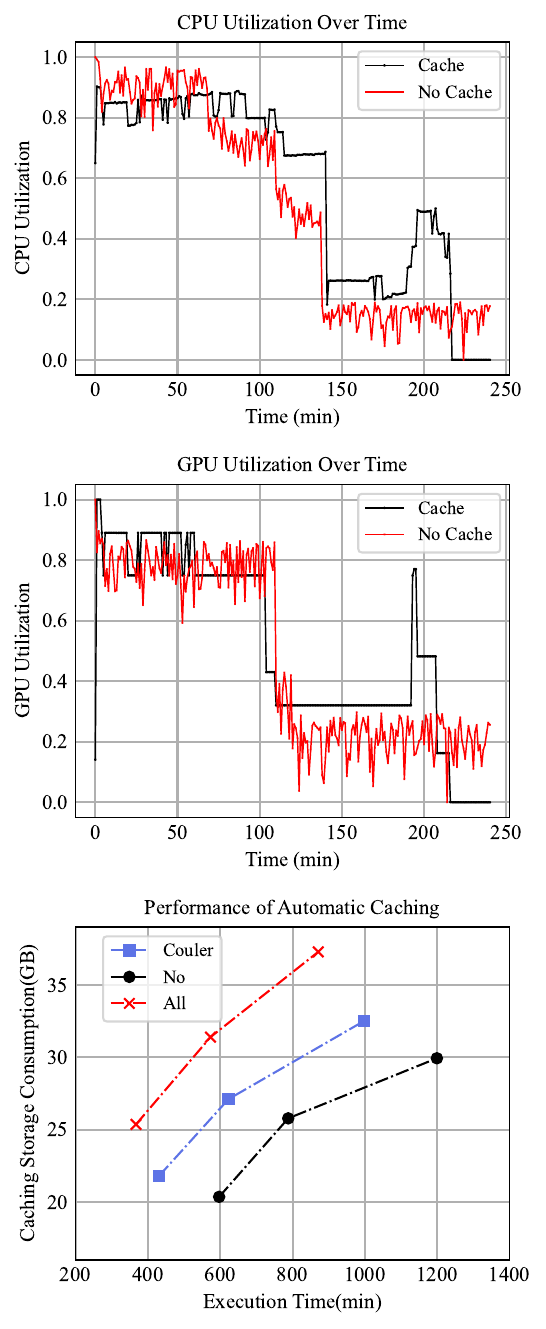}
        \caption{Language Model Fine-tuning}
        \label{fig:language}
    \end{subfigure}
    \caption{Effect of \system on Resource Utils and Workflow Execution Time}
         \vspace{-1.5em}
    \label{fig:cachingchangjing}
\end{figure*}

Initially, we explore three facets of ML workflows: daily usage frequency, typical lifespan, and CPU core usage. We focus on workflows within \company from July 2022 to July 2023. Figure~\ref{fig:workflownumber} illustrates the distribution of the average daily workflow count, revealing a daily average of 22,000 workflows within \company. We define a workflow's lifespan as the hour count between the timestamps of its newest and oldest nodes in its trace, serving as an indicator of its active duration. Figure~\ref{fig:workflowlifetime} depicts that, on average, a workflow within \company remains active for 1 hour. Figure~\ref{fig:workflowcore} presents the average CPU cores utilized by a workflow during its active period, with a mean of 36 cores being used per workflow in \company.

To assess the effectiveness of \system in optimizing workflows, we examine the evolution information of \system within \company from July 2022 to July 2023, as depicted in Figure~\ref{fig:coulerantgroup}. It took approximately ten months to execute all workflows with \system. Figure~\ref{fig:coulercur} reveals that the CPU utilization rate (CUR) in machine learning workflow improved by 18\%. Figure~\ref{fig:coulermur} shows that the memory utilization rate (MUR) improved by 17\%. \system's enhanced fault tolerance significantly improved the workflow completion rate (WCR) for workflows running on 50- and 50+ CPU cores. Due to the high utilization rate of \system, the CUR, MUR, and WCR have seen notable improvements. Therefore, \system has effectively optimized ML workflow performance within \company.

\noindent \textbf{Production insights.} 
Our work is motivated by previous research conducted within Google GCP~\cite{xin2021production}, which highlighted substantial computational waste in ML workflows. Building on these findings, our contributions are diverse, encompassing simplicity and extensibility, automation, efficiency, as well as real-world impact and adoption. We believe the real-world adoption and application of our system by \company and other organizations attest to its practicality and efficacy in production settings. Due to space constraints, we put extensive discussions on production insights into the Appendix E.

\subsection{Performance Study with Caching: RQ2}

\subsubsection{Performance study with Automatic caching}

We evaluate the impact of \system's automatic caching strategy on workflow execution efficiency by comparing execution time and resource utilization against other caching strategies across three different scenarios:
\begin{itemize}
    \item \textbf{Multimodal Training}: This scenario encompasses 37 pods and 19 training models, and involves a training process that integrates various types of input data such as text, images, and sound, aimed at building more robust and adaptable models.
    \item \textbf{Image Segmentation}: This scenario includes 15 pods and 8 training models, focusing on segmenting digital images into multiple parts or sub-regions to identify and locate objects and boundaries within images.
    \item \textbf{Language Model Fine-tuning}: This scenario consists of 21 pods and 11 training models, primarily focusing on further training of pre-trained language models tailored for specific tasks such as text classification or sentiment analysis.
\end{itemize}

We evaluate the execution time and caching storage consumption across five different caching strategies as follows: (1) \textbf{No}, indicating no caching; (2) \textbf{ALL}, involving the caching of all data and intermediate results; (3) \textbf{\system}, representing \system's automatic caching policy. (4) \textbf{FIFO}, first in first out; (5) \textbf{LRU}, least recently used. Based on empirical experience, we choose $\alpha=1.5$ and $\beta=1$ in these experiments for equation~\ref{eq5}.

Figure~\ref{fig:cachingchangjing} illustrates the variations in CPU and GPU usage over time, comparing \system's caching strategy with other caching strategies. Due to space constraints, detailed experimental results for FIFO and LRU can be found in Appendix D.A. 
It is evident that employing \system's caching strategy enhances GPU and CPU utilization, allowing the entire process to complete in less time. 
This is because that automatio cache mechanism can reduce the frequency of I/O operations. And according to existing work~\cite{chien2020tf,pumma2019scalable}, this reduction in I/O overhead is significant as it can substantially decrease the time wasted on these operations, leading to a more efficient workflow execution.
The scatter plot represents the overall execution time and resource consumption of workflows of varying sizes, indicating that \system’s caching strategy achieves higher execution efficiency with a smaller additional resource cost.
\system's strategy tends to conserve resources by avoiding unnecessary caching, yet still reaps the performance benefits of caching the most impactful intermediate results. 
We also calculate the cache hit ratio of the \system caching strategy, which, under reasonably set parameters, averages 84.21\% in production environments, significantly improving the efficiency of workflow execution.

\subsubsection{Performance study with Data caching}
In this section, we investigate the impact of caching on data reading performance by first examining the effect of table caching using two tables from an ads recommendation application, highlighting how caching enhances data loading and deep learning model training efficiency on a hybrid cluster. Secondly, we assess the caching performance for reading small and big files stored remotely, demonstrating significant improvements in data reading speed through local caching. Due to space constraints, we put detailed discussions into the Appendix D.C.

\subsubsection{Performance Study with Cache Sizes}
We further designed experiments to analyze the impact of different cache sizes on \system's performance. In the same three scenarios: Multimodal Training, Image Segmentation, and Language Model Fine-tuning, we set the available cache sizes to 10G, 20G, and a more ample 30G, respectively, and recorded the resource utilization and execution time of the workflows under these conditions. Detailed experimental results can be found in Appendix D.B.
Analysis shows that, under limited cache size conditions, \system can still effectively improve the efficiency of workflow execution, but its effectiveness increases with the size of the cache.

\subsection{NL to Unified Programming Code Generation: RQ3}
\label{nltowexper}
\subsubsection{Experiment Result}

We evaluate the effectiveness of utilizing LLMs to facilitate NL to Unified Programming Code Generation and compare our method with GPT-3.5 and GPT-4, as shown in Table~\ref{tab:codegen}. All models are evaluated at temperatures \( t \in \{0.2, 0.6, 0.8\} \), and we compute pass@k where \( k \in \{1, 3, 5\} \) for each model. The temperature yielding the best-performing pass@k for each \( k \) is selected according to~\cite{nijkamp2022codegen}. 
The 'pass@k' metric is a widely used evaluation method in code generation models. It assesses the model's capacity to produce accurate code within its top 'k' predictions. A higher 'pass@k' percentage indicates the model's reliability in generating correct code options without the need for additional inputs or iterations.
Our method significantly improves the performance of GPT-4 for NL to unified programming code generation and has been widely adopted for \system code generation.

\begin{table}[ht]
\centering
\renewcommand{\arraystretch}{1.4} 
\vspace{-0.5em}
\caption{Evaluation results of our methods with GPT-3.5 and GPT-4. Each pass@k (where \( k \in \{1, 3, 5\} \)) for each model is computed with three sampling temperatures (\( t \in \{0.2, 0.6, 0.8\} \)) and the highest one among the three are displayed, which follows the evaluation procedure in~\cite{nijkamp2022codegen}.}
\begin{tabularx}{0.45\textwidth}{l>{\centering\arraybackslash}X>{\centering\arraybackslash}X>{\centering\arraybackslash}X}
\hline
\multirow{2}{*}{Model} & \multicolumn{3}{c}{pass@\( k \) [\%]} \\
\cline{2-4}
& \( k = 1 \) & \( k = 3 \) & \( k = 5 \) \\
\hline
GPT-3.5 & 35.21 & 37.19 & 39.21 \\
GPT-4 & 45.81 & 48.11 & 50.23 \\
\hline
GPT-3.5 + \textbf{Ours} & 61.25 & 62.97 & 65.03 \\
GPT-4 + \textbf{Ours}  & \textbf{73.12} & \textbf{75.61} & \textbf{77.38} \\
\hline
\end{tabularx}
\label{tab:codegen}
\end{table}

\subsubsection{Running Example}
We provided an example that illustrates the entire process of converting natural language into \system code. It demonstrates the generation of syntactically correct code by the LLM. This example aims to select the best image classification model among ResNet, ViT, and DenseNet by showing the transformation from natural language descriptions to code generation. The details are presented in Appendix C.

\subsubsection{Cost Analysis and Future Prospects Discussion}
On the matter of cost, we understand the concerns about the economic feasibility of deploying LLMs, especially considering the cost per token for each workflow. So we present the average costs for each workflow in terms of the number of tokens processed by LLMs and the corresponding money for model “GPT-3.5-turbo” and “GPT-4” in Table~\ref{tab:cost-analysis}. \system has shown promising results for a range of tasks except for some complex workflows. The use of LLMs in our research is primarily aimed at exploring the potential of these models to streamline and enhance the code generation process.

\begin{table}[ht]
\centering
\caption{Cost Analysis of Workflow Generation}
\label{tab:cost-analysis}
\begin{tabular}{ccc}
\toprule
\multirow{2}{*}{\textbf{Cost / Workflow}} & \multicolumn{2}{c}{\textbf{Workflow Generation}}  \\
\cmidrule(lr){2-3} 
            & GPT-3.5-turbo & GPT-4  \\
\midrule
Token       & 3212.1    & 3813.7  \\
\midrule
Money (\$)  & 0.005     & 0.140     \\
\bottomrule
\end{tabular}
\end{table}

\vspace{-0.5em}

We acknowledge that current LLMs already demonstrate a satisfactory level of accuracy in code generation. Indeed, we are considering fine-tuning as a viable method to enhance the quality of the generated code. Specifically, our team has conducted work on Multi-LoRA optimization for fine-tuning~\cite{multilora}. Additionally, we are developing a workflow for code generation by fine-tuning `llama2'. We will soon release a fine-tuned model in the \system open-source repository.

\begin{figure}[ht]
    \centering
    \begin{subfigure}{0.24\textwidth}
        \includegraphics[width=\linewidth]{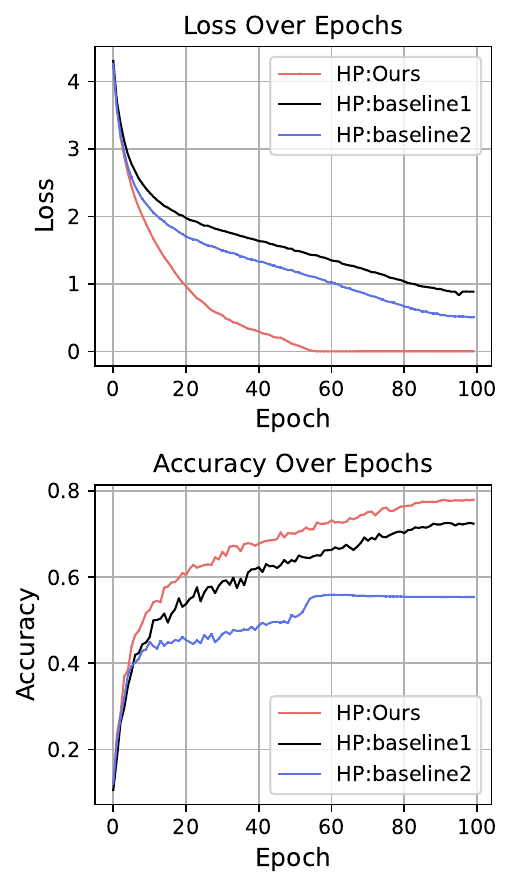}
        \caption{Auto Configuration for CV}
        \label{fig:autoconcv}
    \end{subfigure}
    \hfill
    \begin{subfigure}{0.24\textwidth}
        \includegraphics[width=\linewidth]{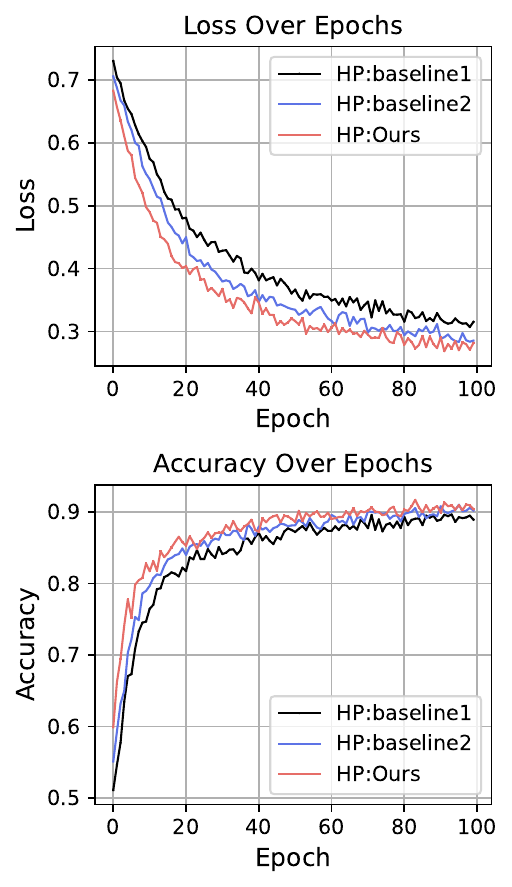}
        \caption{Auto Configuration for NLP}
        \label{fig:autoconnlp}
    \end{subfigure}
    \caption{Effect of Auto Hyperparameter Configuration}
    \label{fig:autochaocan}
\vspace{-5mm}
\end{figure}

\subsection{Automatic Hyperparameter Configuration: RQ4}

We next evaluate the performance of automatic hyperparameters configuration using LLMs to generate recommended hyperparameters.
Following the workflow detailed in Workload, we apply automated hyperparameter tuning to both the \textit{cv} and \textit{nlp} modules, with \textit{HP:Ours} as our recommended parameter. As depicted in Figure~\ref{fig:autochaocan}, the recommended parameters exhibit the lowest loss and the highest accuracy, showcasing the potent capability of \system's Automatic Hyperparameter Tuning. 
HP-baseline1 represents a set of hyperparameters that were manually selected based on expert knowledge and empirical best practices in the field. HP-baseline2 corresponds to a set of hyperparameters derived from historical benchmarks and recommendations in the literature.

\vspace{-1em}
\subsection{Comparative Learning Analysis of Workflow Engines}

To assess the effectiveness of our system's unified programming model, we conducted a survey, where we asked 15 engineers who were not familiar with these workflow engines to learn and use code snippets with similar functionality in \system, Argo, and Airflow. We measured the time it took for them to understand and work with the provided code samples. The results of this survey in Table~\ref{tab:learning-analysis} clearly indicate that our \system API is more user-friendly and easier to learn.

\begin{table}[ht]
\centering
\caption{Workflow Learning Comparative Analysis}
\label{tab:learning-analysis}
\begin{tabular}{cccc}
\toprule
\multirow{2}{*}{\textbf{Time / Workflow}} & \multicolumn{3}{c}{\textbf{Workflow Engines}}  \\
\cmidrule(lr){2-4} 
            & \system & Argo  & Airflow \\
\midrule
Time (min)      & 18   & 61    & 50 \\
\bottomrule
\end{tabular}
\end{table}

\vspace{-5mm}

\section{Related work}
\label{sec:related_work}

\textbf{Jobs Scheduling in the Cloud.}
One machine learning workflow usually includes different stages to produce the model, and each stage/step is associated with different kinds of jobs~\cite{kubeflow,mlflow,survey_ml,sang2023cougar}. Kubernetes~\cite{k8s} boasts a rapidly growing community and ecosystem, providing robust support for workflows. Kubernetes is based on a highly modular architecture that abstracts the underlying infrastructure and allows internal customization. It supports various big-data frameworks (e.g. Apache Hadoop MapReduce~\cite{hadoop-hdfs, hadoop}, Apache Spark~\cite{spark-nsdi}, Apache Kafka~\cite{Kafka}, Apache Flink~\cite{flink} etc). More recently, Kubeflow~\cite{kubeflow} allows users to submit distributed machine learning tasks on Kubernetes.

\textbf{Workflow for AI/Machine Learning.}
One machine learning workflow usually includes different stages to produce the model, and each stage/step is associated with different kinds of jobs~\cite{kubeflow,mlflow,survey_ml}. Kubernetes~\cite{k8s} boasts a rapidly growing community and ecosystem, providing robust support for workflows. TFX~\cite{tfx} is a TensorFlow based AI framework for machine learning model training, but it is specifically designed for TensorFlow only. Some works in the HCI community study ML/DS workflows by interviewing ML developers and data scientists~\cite{zhang2020data}.

\textbf{Workflow Engine.}
A workflow engine is a software application that manages business processes. 
Argo Workflows ~\cite{Argo} is an open-source container-native workflow engine for orchestrating parallel jobs on Kubernetes. Apache Airflow~\cite{Airflow} is a Python-based platform for running directed acyclic graphs (DAGs) of tasks. Apache OOzie~\cite{ooize} is a workflow engine in the Hadoop ecosystem. Kubeflow~\cite{kubeflow} has a sub-project called Kubeflow Pipelines for end-users to develop machine learning pipelines. The complementary list of workflow engines can be found in link ~\cite{workfloweng}. There are also recently workflow engines like Ray~\cite{moritz2018ray}, CodeFlare~\cite{codeflare} and ThunderML~\cite{shrivastava2019thunderml}. 
\system is inspired by the design of PyTorch~\cite{pytorch},
which compiles a high-level AI model to a DAG.

\textbf{Automated Machine Learning.}
Automated Machine Learning (AutoML) \cite{hutter2019automated, truong2019towards} simplifies the process of machine learning model selection and hyper-parameter tuning, thereby making ML more accessible to non-experts.
In the last decade, substantial advancements in AutoML have emerged with the introduction of open-source frameworks such as Auto-WEKA \cite{kotthoff2019auto, thornton2013auto}, AutoSklearn \cite{feurer2015efficient}, AutoGluon \cite{erickson2020autogluon}, and Auto-PyTorch \cite{zimmer2021auto}, alongside commercialized frameworks.

\textbf{Workflow Optimization and Query Optimization.}
Workflow optimization applies broadly, covering areas such as scientific workflows, business processes, and cloud computing, and addresses tasks beyond data processing, including computational and data movement activities. It also faces unique challenges like deadlines, budget constraints, and fault tolerance, which are less common in query optimization. Additionally, workflow optimization utilizes specific strategies like dynamic scheduling, partitioning, and machine learning for predictive optimization, highlighting its distinct requirements compared to the more static nature of query optimization.

\pdfoutput=1
\section{Conclusion}
\label{sec:conclusion}

In this paper, we introduced \system, a system designed for unified machine learning workflow optimization in the cloud. \system simplifies ML workflow generation using NL descriptions, abstracting the complexities associated with different workflow engines. 
Furthermore, \system boosts computational efficiency through automated caching, large workflow auto-parallelization, and hyperparameter tuning.

\pdfoutput=1
\UseRawInputEncoding
\section{Acknowledgement}

We want to thank Wei Yan who led the initial discussions of the design and implementation of this project. We would like to thank many colleagues at Alibaba Group and Ant Group for their valuable advice that improved this project, namely, Lan Li, Jian Liu, Wei Shu, Changhua He, Yi Wang, Jun Jiang, Chunbo Dai, Chenyi Ding, Mu Xiong, Yitao Shen, Yi Zhang, Xiaohua Cai, and Qinglong Wang. 
In addition, we appreciate the timely review and feedback from the Argo team every time when we contributed bug fixes and new features to not only meet our internal requirements but benefit the community users. 
Last but not least, we are grateful for the contributions from the \system open source community and their initial public adoption and feedback that improves the project. 


\nocite{alipayio}
\nocite{alluxiopart1}
\nocite{tfDarshan}
\nocite{pumma2019scalable}
\nocite{chien2020tf}
\nocite{multilora}
\nocite{xin2021production}
\nocite{argo2couler}
\nocite{AdoptersofCouler}

\bibliographystyle{IEEEtranS}
{\footnotesize 
\bibliography{ieeexample}}

\begin{appendices}
\section{Unified programming model}
\label{sec:interface}

This section provides an overview of the programming model to define a workflow. We at first introduce the programming interface, then show some examples for running a workflow in a machine learning application. The major design rule of \system is aiming to help end-users to write a workflow without the specific knowledge of the workflow engine itself. In this paper, we provide two ways to define a workflow. One way is building a workflow implicitly (e.g., code~\ref{lst:sequence_example} and code~\ref{lst:control_example}), the other way is explicitly defining a workflow as code~\ref{lst:dag_example}. 
The main difference is whether DAG is described in the program. 
The core functions of \system are listed in Table~\ref{tab:api_summary}, we would illustrate how to use \system to build a workflow based on the following examples. Because most data scientists prefer to use Python, the programming interface of \system is based on Python. However, this design is not limited to Python and could be extended to Java or another language. At Ant Group, we also provided Java client for end-users. 

\begin{table}[h]
\begin{center}
\scalebox{0.8} {
\begin{tabular}{ | l | l | l| l | l| l|}
    \hline
    Name 
    & API 
    & Description
    \\ \hline
     Run script & $couler.run\_script()$ & Run a script in a Pod  \\ \hline
     Run container & $couler.run\_container()$ & Start a container   \\    \hline
     Run job & $couler.run\_job()$ & Start a distributed job  \\    \hline
   	 Condition & $couler.when()$ & Condition definition  \\    \hline
   	 Map & $couler.map()$ & Start multiple instances for one job  \\    \hline
   	 Concurrent & $couler.concurrent()$ & Run multiple jobs at the same time  \\ \hline
     Recursive  & $couler.exec\_while()$ & Run a function until a condition meets \\ 
    \hline
  \end{tabular}
}
\end{center}
\caption{API Summary of \system}
\vspace{1em}
\label{tab:api_summary}
\vspace{-1em}
\end{table}

\begin{lstlisting}[caption={Basic workflow and artifact definition in \system}, label={lst:sequence_example}, frame=bt]
def producer(step_name):
    output_path = "/opt/hello_world.txt"
    output_place = 
        couler.create_parameter_artifact(
         path=output_path, is_global=True
    )
    return couler.run_container(
        image="docker/whalesay:latest",
        args=["echo -n hello world > 
        %s" % output_place.path],
        command=["bash", "-c"],
        output=output_place,
        step_name=step_name,
    )

def consumer(step_name, input):
    couler.run_container(
        image="docker/whalesay:latest",
        command=["cowsay"],
        step_name=step_name,
    )

output = producer("step1")
consumer("step2", output)
\end{lstlisting}

\subsection{Basic workflow example and artifact}

A workflow is made by different steps, then each step is isolated from each other in the cloud base on the container. A container manages the complete the lifecycle of its host system, the contained environment helps each step to own the specific computing requirement and resource. However, this brings issues to pass data from one step to the following step in a workflow. In this work, we introduce the artifact to help users to store the intermediate results inside a workflow. 

An artifact is a by-product of workflow development and created. This might include things like data set, parameter, and diagram, etc. For example, a machine learning pipeline generates statistic results, trained models, or new features. For different kinds of artifacts, users can register different physical storage to place the related artifact based on specific requirements. Thus, we introduce multiple ways to store artifacts as Table~\ref{tab:artifact}. 

Take the code~\ref{lst:sequence_example} as a running example, users register a parameter artifact to pass data among two steps (e.g., a producer and consumer). Function $producer()$ generate a message and pass the message to function $consumer()$ from line~1 to line~15. Each function is built based on $couler.run\_container()$, where $couler.run\_container()$ is used to start a Pod in Kubernetes and run the corresponding function. As a result, two Pods start and run step by step in the workflow as running a local python code lines~22 to 23. Then, the workflow engine propagates the related value among steps without users' interaction. Note, Pods are the smallest deployable units of computing that you can create and manage in Kubernetes.

\begin{table}[tp]
\begin{center}
\scalebox{0.80} { 
 \begin{tabular}{ | l | l | l| l | l| l|}
    \hline
    Name 
    & API 
    & Description
    \\ \hline
     Parameter & $couler.create\_parameter\_artifact$ & Create a parameter  \\ \hline
     HDFS & $couler.create\_hdfs\_artifact$ & Create a HDFS artifact   \\    \hline
     Amazon S3 & $couler.create\_s3\_artifact$ & Create a S3 artifact   \\    \hline
   	 Alibaba OSS & $couler.create\_oss\_artifact$ & Create a OSS artifact   \\    \hline
   	 Google GCS & $couler.create\_gcs\_artifact$ & Create a GCS artifact    \\    \hline
   	 Git storage & $couler.create\_git\_artifact$ & Create a Git artifact   \\    \hline
  \end{tabular}
}
\end{center}
\caption{Artifact support in \system}
\vspace{1em}
\label{tab:artifact}
\end{table}

\subsection{Control flow}

\begin{lstlisting}[caption={Workflow control in \system}, label={lst:control_example}, frame=bt]
def random_code():
    import random

    res = "heads" if random.randint(0, 1) 
        == 0 
    else "tails"
    print(res)

def flip_coin():
    return couler.run_script(
        image="python:alpine3.6", 
        source=random_code)

def heads():
    return couler.run_container(
        image="alpine:3.6", 
        command=["sh", "-c", 
        'echo "it was headed"']
    )

def tails():
    return couler.run_container(
        image="alpine:3.6", 
        command=["sh", "-c", 
        'echo "it was tailed"']
    )

result = flip_coin()
couler.when(couler.equal(result, "heads"), 
  lambda: heads())
couler.when(couler.equal(result, "tails"), 
  lambda: tails())
\end{lstlisting}

Control flow is important for defining a workflow. For a machine learning workflow, if a trained model fails to meet the online serving criteria, the following step (e.g., model deployment step) would not push the model to a server rather than alerting user the model training failures. 

This example code~\ref{lst:control_example} combines the use of a Python function result (e.g., Function $random\_code()$), along with conditionals, to take a dynamic path in the workflow. In this example, depending on the result of the first step defined in $flip\_coin$ (line~27) , the following step will either run the $heads()$ step (line~28) or the $tails()$ step (line~29). We can notice, steps in \system can be defined via either Python functions or script to running for containers (e.g., line~9). In addition, the conditional logic to decide whether to flip the coin in this example is defined via the combined use of $couler.when()$ and $couler.equal()$. As a result, users can control the workflow logic based on the results of steps in the workflow dynamically. 

\subsection{Define a workflow explicitly}
\begin{lstlisting}[caption={Workflow DAG in \system}, label={lst:dag_example}, frame=bt]
def job(name):
    couler.run_container(
        image="docker/whalesay:latest",
        command=["cowsay"],
        args=[name],
        step_name=name,
    )

#   A
#  / \
# B   C
#  \ /
#   D
def diamond():
    couler.dag(
        [
            [lambda: job(name="A")],
            [lambda: job(name="A"), 
                lambda: job(name="B")],  # A -> B
            [lambda: job(name="A"), 
                lambda: job(name="C")],  # A -> C
            [lambda: job(name="B"), 
                lambda: job(name="D")],  # B -> D
            [lambda: job(name="C"), 
                lambda: job(name="D")],  # C -> D
        ]
    )
diamond()
\end{lstlisting}

In general, data scientists prefer to build a workflow implicitly as the examples mentioned above. On other hand, data engineers want to organize a workflow explicitly. \system support end-users to build the dependency among steps based on function $set\_dependencies$. Function $set\_dependencies$ take input as a function and let the user to define the dependencies steps of others based on the step name. For example, the code~\ref{lst:dag_example} generates a diamond workflow as line~15. In this way, users need to own a clear big picture for the workflow, and under how the running logic among steps in their real application. At Ant Group, we analyze the users' preference to choose to build a workflow and we found major of data scientists choose to define a workflow implicitly, yet, data engineers incline to build a workflow explicitly since they facing more than one hundred steps in a workflow. The definition of DAG workflow via explicit way helps data engineer to debug a failed workflow more easily, and build a complicated workflow with hundred nodes. 

\subsection{Example: running recursive in a workflow}
\system provide a straightforward way to help end users to define the recursive logic in a workflow. For machine learning workflow, data scientists need to search a machine learning model until the model meets the specific requirement (e.g., model precision, convergence ratio, or the number of iteration steps is bigger than predefined value). Thus, data scientist hope to search the best ML model in a workflow recursively. 
Example~\ref{lst:recursive_example} demonstrates how to run the recursive in a workflow. This $flip\_coin()$ step is running recursively until the output is equal a $tails$ (line 14 to 15). 

\begin{lstlisting}[caption={Recursive in \system}, label={lst:recursive_example}, frame=bt]
def random_code():
    import random

    result = "heads" if random.randint(0, 1)
        == 0 else "tails"
    print(result)


def flip_coin():
    return couler.run_script(
        image="alpine3.6",
        source=random_code)

# Stop flipping coin until the outputs of 
    'flip_coin' is not 'tails' 
couler.exec_while(couler.equal("tails"), 
    lambda: flip_coin())
\end{lstlisting}

\subsection{Example: select a best ML model}

\begin{lstlisting}[caption={Searching a best ML model in \system}, label={lst:map_example}, frame=bt]
def train_tensorflow(batch_size):
    import couler.steps.tensorflow as tf

    return tf.train(
        num_ps=1,
        num_workers=1,
        command="python /train_model.py",
        image="wide-deep-model:v1.0",
        input_batch_size=batch_size,
    )


def run_multiple_jobs(num_jobs):
    para = []
    i = 0
    batch_size = 0
    while i < num_jobs:
        batch_size += 100 
        para.append(batch_size)
        i = i + 1

    return couler.map(lambda x: 
        train_tensorflow(x), para)
        
def evaluation(model_path):
    return couler.run_container(
        image="model_evalutation:v1",
        command=["python model_eval.py"],
        args=[model_path],
        step_name="eval",
    )

model_path = run_multiple_jobs(5)
couler.map(lambda x: evaluation(x), 
    model_path)
\end{lstlisting}

The hyper-parameters of machine learning modes such
as batch size or converge ratio decide the performance of the model. Data scientists need to run multiple jobs in one same workflow to find the best model based on the same input data. 

The sample program ~\ref{lst:map_example} implements a model searching procedure for a deep learning model (e.g., wide and deep model ~\cite{cheng2016wide}. This is a common recommendation algorithm
that recommends items to users based on users' profiles and user-item interaction. We start by defining a training job via the step zoo of~\system, this job train a DL model based on the different batch size from line~1 to 10, then run $map$ function to start multiple TensorFlow jobs in the same workflow from line 13 to 33. Next, multiple evaluation steps are running based on the outputs of previous model training results from line 25 to 31.

\subsection{Example: Running an AutoML pipeline}

\begin{lstlisting}[caption={An AutoML workflow in \system}, label={lst:concurrent_example}, frame=bt]
def train_xgboost():
     train_data = Dataset(
        table_name="pai_telco_demo_data",
        feature_cols="tenure,age,
            marital,address,ed,employ",
        label_col="churn",
    )

    model_params = {"objective": 
        "binary:logistic"}
    train_params = {"num_boost_round": 10, 
        "max_depth": 5}

    return xgboost.train(
        datasource=train_data,
        model_params=model_params,
        train_params=train_params,
        image="xgboost-image",
    )

def train_lgbm():
    train_data = Dataset(
        table_name="pai_telco_demo_data",
        feature_cols="tenure,age,
            marital,address,ed,employ",
        label_col="churn",
    )
    
    lgb = LightGBMEstimator() 
    lgb.set_hyperparameters(num_leaves=63, 
        num_iterations=200)
    lgb.model_path = "lightgbm_model"
    return lgb.fit(train_data)
    
couler.concurrent([lambda: train_xgboost(), 
    lambda: train_lgbm()])
\end{lstlisting}

A more complex machine learning workflow application is the AutoML. Different from hyper-parameter tuning, data scientist prefers to select the best models from multiple model candidates based on the same input data. This program ~\ref{lst:concurrent_example} shows how to choose a best model from two machine learning model (e.g., XGBoost~\cite{XGBoost} and LightGBM~\cite{lgbm}), which are the state-of-art tree and boost based machine learning model. The training model in $train\_xgboost()$ and $train\_lightbm()$ is defined based on user's from line 1 to line 33, then $couler.concurrent()$  will run two jobs parallel in the same workflow. Different from the way of $couler.map()$, Function $couler.concurrent()$ start two training process based on different machine learning model.

\section{Implementation}
\label{sec:implementation}

The Python SDK for \system is now open-source, as shown in its public repository \footnote{\url{https://couler-proj.github.io/couler/}}. Several top enterprises have integrated this SDK into their production environments. The whole \system service is crafted in Golang, encompassing all internal components. Initially, the service might remain proprietary due to user onboarding challenges, but enhancing and ensuring the reusability of the optimization components is our priority. Thus, we are developing these components as core libraries. Consequently, the service operates as a gRPC service atop these libraries. In its open-source form, the Python SDK can utilize these libraries for extended capabilities. 

It's worth mentioning that \system is extensively used by Ant Group, managing over 20,000 workflows and 250,000 pods daily. Insights from our deployment experiences with the workflow engine are discussed in subsequent sections.

Beyond AntGroup, COULER is being used by a diverse set of organizations, as indicated in the list of adopters available on Adopters of Couler~\cite{AdoptersofCouler}. These adopters span different industries and use cases, suggesting that COULER is flexible and adaptable to various requirements and scenarios. Moreover, Argo's endorsement of Couler ~\cite{argo2couler} further supports COULER's suitability for different workloads and platforms.

\subsection{Workflow scheduling among clusters}
At Ant Group, we have more than one clusters in different locations. Workflows are scheduled among those clusters and each cluster has its specific configuration. For example, Cluster A is specifically designed for GPU jobs, Cluster B is located far away from the storage cluster, Cluster C provides more CPU capacity than others. In addition, the storage and computation capacity of a cluster is changing with respect the time. We need to make sure each cluster owns the similar computation load. In this work, we provide a workflow queue to schedule the related steps of workflow into a corresponding cluster based on the following properties: (a) the workflow priority based on business logic, (b) cluster current capacity of CPU/Memory, (c) user's current CPU/Memory quota, (d) user's current GPU quota. Then, a job is queued and pull out from the queue based on the weight combination of mentioned factors. In this way, we can guarantee each cluster shares a similar capacity and avoid one cluster being overflow in the production. 

\subsection{Monitor and failure handler}
In order to reduce the unnecessary failure of workflow belonging  to the system environment (i.e., abnormal patterns of cloud), we adopt following polices to improve the stability: (a) workflow on-time monitor, (b) workflow controller auto retry, (c) provide options for users to restart from failure.  

Initially, we monitor workflow status and track the health status of the workflow engine. For example, we record the number of workflows based on their status, the latency for the workflow operator to process a workflow, etc. This monitor metric helps the SRE to respond to the abnormal behaviors of the workflow at the first time. 

Subsequently, we get the patterns of  system errors related workflow. For example, ``ExceededQuotaErr" means the Etcd of Kubernetes exceeded quota during the system is updating. ``TooManyRequestsErr" means too many requests being handled by API-server, usually happens under high pressure. The backoff limit retry policy would help avoid DDOS of the cluster Etcd server. In general, we have found more than 20 abnormal patterns to retry, then the workflow controller restarts the failed step inside a workflow rather than from the begging automatically.

Furthermore, there are instances where users prefer to manually retry some failed workflows, a scenario frequently encountered in machine learning. In such cases, data scientists update the relevant steps and wish to retry the workflow from the failure point instead of from the beginning. To address this type of failure, \system's server first retrieves the failed workflow from the database. Note that we persist workflow metadata into a database for automated management. The server then processes the failed workflow, skipping the steps with ``Succeeded," ``Skipped," or ``Cached" status. Subsequently, the server deletes the failed steps and the related CRDs and marks these steps as running. Finally, the workflow's status is updated to running, and it is restarted from the failed step by the workflow operator.

\subsection{Caching input data for machine learning workflow}

In a machine learning workflow, the input data for model training is stored in a data storage cluster, while the machine learning job runs in a separate computation cluster. This necessitates fetching data from remote storage before training, which is time-consuming and can lead to network IO failures. This is particularly problematic for applications like ads recommendation, where input tables often exceed 1TB, and for image and video deep learning models, which usually involve over a million files.

An analysis of production machine learning workflows at Ant Group, which include more than 5k models and 10k workflows for applications such as ads recommendation, fault detection, and video and image analysis, revealed that most workflows read the same data multiple times. For instance, 70\% and 85\% of the input data for tables and files, respectively, was read repeatedly. This redundancy arises due to several factors: (1) a single training job may need to scan the entire dataset multiple epochs, (2) multiple training jobs may need to read the same data to train the basic model, and (3) different training jobs may read overlapping data partitions using a sliding time window.

Currently, users read input data via Python/Java clients in the Kubernetes pods of machine learning training jobs. However, the workflow engine, such as Argo Workflows, cannot track data flow because it operates on Kubernetes Custom Resource Definitions (CRD) and does not monitor the runtime information of pods. This leads to two issues: (1) if a workflow fails, the training job must read the input data again, and (2) if multiple training jobs in the same workflow read the same input data, each job reads the data remotely, leading to redundant data access and high network IO.

To address these issues, we propose a new Kubernetes CRD, called \textit{Dataset}, to represent the input and output data of a job. The schema of \textit{Dataset} is shown in Code~\ref{lst:dataset_example}. This CRD records the metadata of the data, enabling the workflow engine to understand the input and output of a training job and skip steps to read cached data. Additionally, a caching server reads the \textit{Dataset} status and syncs the data from the storage cluster to the computation cluster, eliminating the need for multiple data synchronizations for different jobs.

\begin{lstlisting}[caption={Dataset CRD in \system}, label={lst:dataset_example}, frame=bt]
apiVersion: io.kubemaker.alipay.com/v1alpha1
kind: Dataset
metadata:
  name: couler-cache-dataset
spec:
  owner: user_id
  odps:
    accessID:
      secretKeyRef:
        name: test-dataset-secret
        key: aid
    accessKey:
      secretKeyRef:
        name: test-dataset-secret
        key: akey
    project: test_project
    table: test_table
\end{lstlisting}
\vspace{1em}

\subsection{Interactive GUI}
\label{implementation:gui}
 
\begin{figure}[h]
\centering
\includegraphics[width=1\columnwidth]{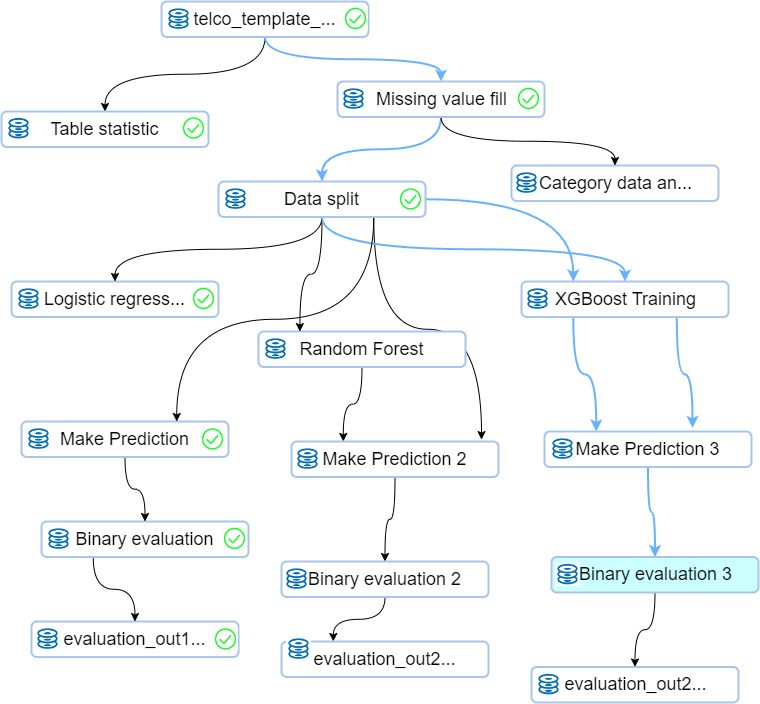}
\caption{GUI for the workflow}
\vspace{0.5em}
\label{fig:workflow_gui}
\end{figure}

In addition to the programming API for defining workflows, we also offer a GUI interface within a web portal. With this approach, users can create workflows without any programming experience. Let's consider Figure~\ref{fig:workflow_gui} as an example. Users aim to identify the best model for predicting user churn. Data scientists initially define data splitting methods for training, select various well-known models (e.g., logistic regression, random forest, and XGBoost) for training the same data, and ultimately choose the best model based on evaluation results. End-users only need to configure model-related parameters or data splitting methods. The backend then translates these actions into the workflow's IR, as explained in Section~\ref{sec:system_overview}, which is subsequently sent to the server for further optimization.

Meanwhile, machine learning algorithm developers can construct their own models and share them with others on the same platform. This collection of well-known machine learning algorithms is referred to as the "model zoo." A model zoo comprises model definitions and trained model parameters, essential for using the model in predictions and other analytical tasks. Notably, the backend of the model zoo corresponds to the "step zoo" of \system, as each model runs as one step in a workflow. Therefore, the GUI and related actions align with \system's programming. Leveraging the interactive GUI, a significant portion of workflows (e.g., over 60\%) in the cluster are executed via the GUI, addressing the rapid development needs of machine learning applications.

\subsection{SQL and SQLFlow}
\label{implementation:sql}

In addition to the GUI and Python programming interface, SQLFlow~\cite{wang2020sqlflow} offers an SQL-like language to train machine learning models and employ the trained models for predictions. \system serves as the default backend for SQLFlow~\cite{wang2020sqlflow}. All optimizations discussed in this work aim to enhance SQLFlow's model training speed. Typically, a SQLFlow SQL statement is converted into Couler programming code, which then initiates a workflow in Kubernetes. An example can be seen in code~\ref{example:sql_flow_train}, where a DNNClassifier model is trained using TensorFlow Estimators~\cite{tfestimators} on the sample data, Iris.train.

\vspace{0.5em}

\lstset{
  language=SQL, 
  basicstyle=\ttfamily, 
  keywordstyle=\color{blue},
  frame=none
}

\begin{lstlisting}[label=example:sql_flow_train]
SELECT *
FROM iris.train
TO TRAIN DNNClassifier
WITH model.n_classes = 3, 
model.hidden_units = [10]
COLUMN sepal_len, sepal_width, 
    petal_length
LABEL class
INTO sqlflow_models.my_dnn_model;
\end{lstlisting}

\vspace{0.5em}

Based on the trained model above, the user can submit a SQL query to get the validation data, then apply the trained model to make a prediction~\ref{example:sql_flow_predict} over the selected data. The output of SQL is the data with the prediction value. Naturally, users also could start a data analysis job over the predicted results based on SQL. 

\vspace{0.5em}

\begin{lstlisting}[label=example:sql_flow_predict]
SELECT *
FROM iris.test
TO PREDICT iris.predict.class
USING sqlflow_models.my_dnn_model;
\end{lstlisting}
\
\section{Running Example of NL to Unified Programming Interface}

\begin{figure}
\centering
\includegraphics[width=0.45\textwidth]{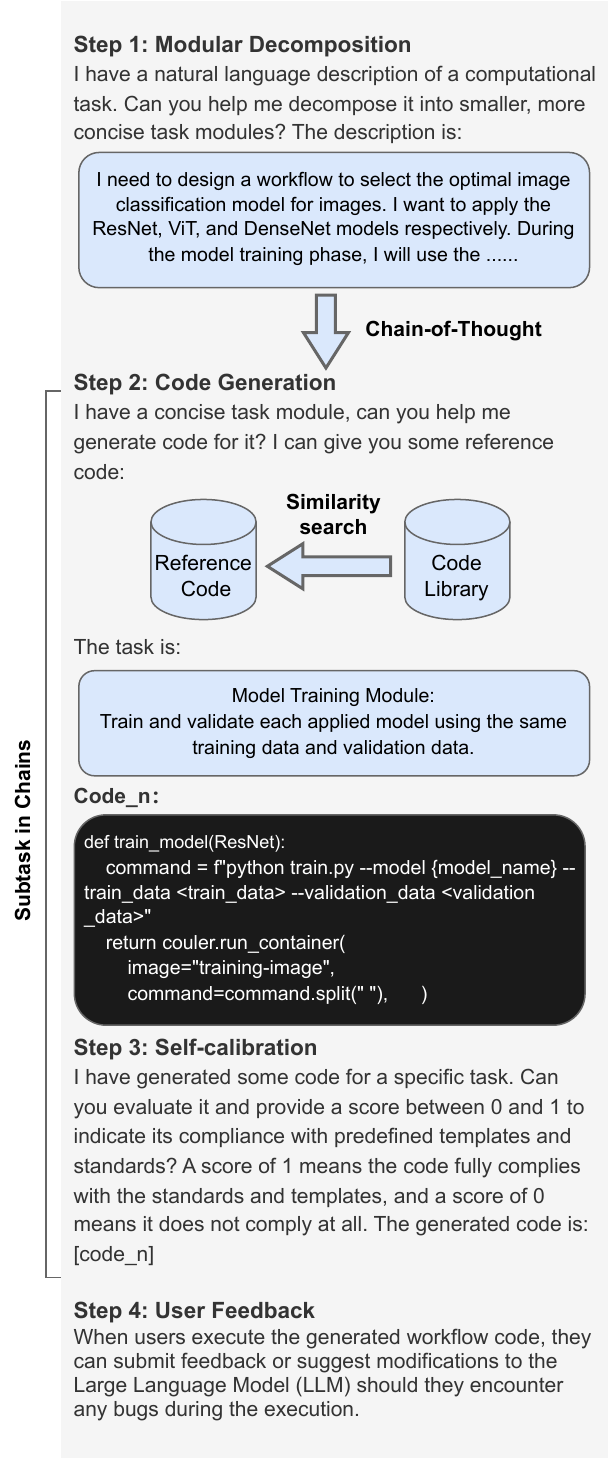}
\caption{Running Example: NL to Unified Programming Code}
\vspace{0.5em}
\label{fig:automlexample}
\end{figure}

We illustrated an example in Figure~\ref{fig:automlexample}, displaying the full automated process of converting Natural Language to Unified Programming Coded. The goal of this example is to choose the optimal image classification model from ResNet, ViT, and DenseNet, by showcasing the transformation from natural language descriptions to code generation.

\noindent \textbf{Step 1: Modular Decomposition}

Initially, we employed a \textit{chain-of-thought} strategy to break down the original natural language descriptions into smaller, more concise task modules. For the given workflow description: \textit{I need to design a workflow to select the optimal image classification model...}. Through this strategy, we identified the following task modules: Data Loading, Model Application (ResNet, ViT, DenseNet), Model Training, Model Validation, Model Comparison, and Model Selection.

\noindent \textbf{Step 2: Code Generation}

For each independent task module, we leveraged Large Language Models (LLMs) to generate code. Given each module has a clear and singular task, this enhances the accuracy and reliability of the generated code. For instance, for the \textit{Model Training} task, we generated the relevant code related to training models, ensuring all models use the same training and validation datasets.

\noindent \textbf{Step 3: Self-calibration}

Subsequently, we incorporated a self-calibration strategy to optimize the generated code. This strategy offers improvement suggestions by comparing the generated code with predefined templates in terms of similarity, and these suggestions can be automatically applied to refine the code further, ensuring its compliance with \system norms and standards. For example, the generated code for model training was compared and optimized against a predefined training code template as necessary.

\noindent \textbf{Step 4: User Feedback}

Finally, users have the opportunity to review and validate the generated \system code. If it does not meet users' requirements, they can provide feedback and suggestions. The system will utilize this feedback to optimize the generated code, enhancing the precision of code generation in future tasks. For instance, if users find the model comparison methodology to be insufficient or biased, they can suggest modifications, allowing the system to adjust the code accordingly based on user feedback.

Through these steps, we not only transformed natural language descriptions into executable code but also ensured the precision, consistency, and efficiency of the generated code. This enables users with limited programming experience to easily realize their computational tasks and workflow needs.

\section{Cache Strategy and Cache Size Ablation Experiments}

This section showcases the comparative effects of the \system caching strategy against the FIFO and LRU caching strategies. Additionally, this section displays the performance of the \system caching strategy under different cache sizes. The following experimental graphs each show the differences in execution time and CPU/GPU utilization for various workflow tasks under different conditions.

\subsection{Performance Study with Automatic Caching}

FIFO and LRU, as two efficient and universal caching strategies, can effectively improve the cache hit rate and resource utilization efficiency during the caching process. This paper compares the \system caching strategy with these two strategies, conducting tests in three scenarios: Image Segmentation, Language Model Fine-tuning, and Multimodal Training. The results indicate that compared to the FIFO and LRU strategies, the \system caching strategy is more adaptable to the production environment of workflows. By considering artifact reconstruction cost, artifact reuse value, and artifact caching cost, \system achieves higher production efficiency.

\label{sec:exper}
\begin{figure}[ht]
    \centering
    \begin{subfigure}[b]{0.24\textwidth}  
        \includegraphics[width=\textwidth]{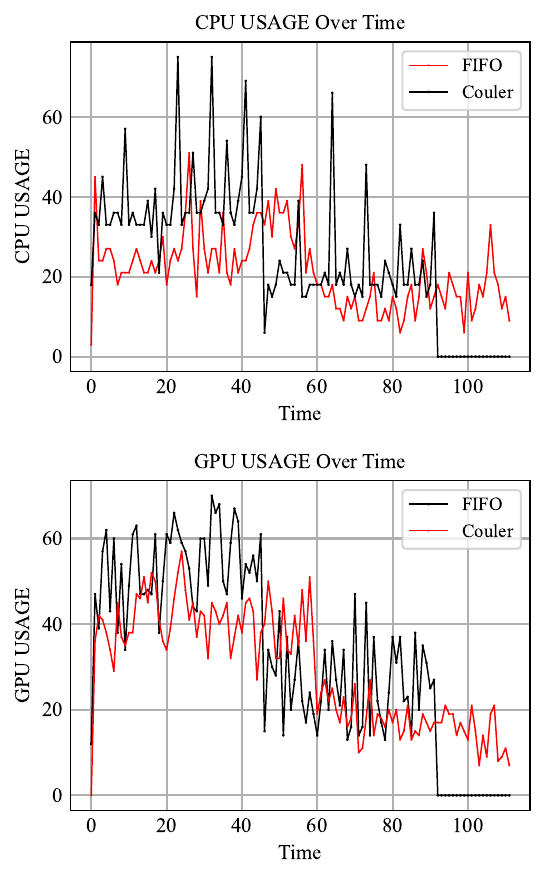}
        \caption{FIFO}
        \label{fig:Figure_CV_FIFO}
    \end{subfigure}
    \hfill  
    \begin{subfigure}[b]{0.24\textwidth}  
        \includegraphics[width=\textwidth]{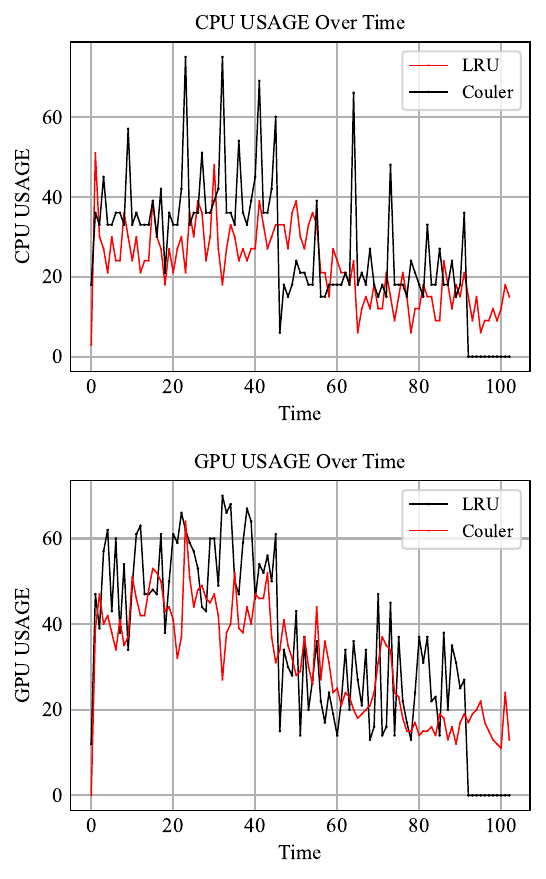}
        \caption{LRU}
        \label{fig:Figure_CV_LRU}
    \end{subfigure}
    \caption{
Effect of \system on Resource Utilization and Workflow Execution Time in Image Segmentation Scenarios with FIFO and LRU Strategies}
    \label{fig:cache_perf1}
    \vspace{-1.0em}
\end{figure}

\begin{figure}[ht]
    \centering
    \begin{subfigure}[b]{0.24\textwidth}  
        \includegraphics[width=\textwidth]{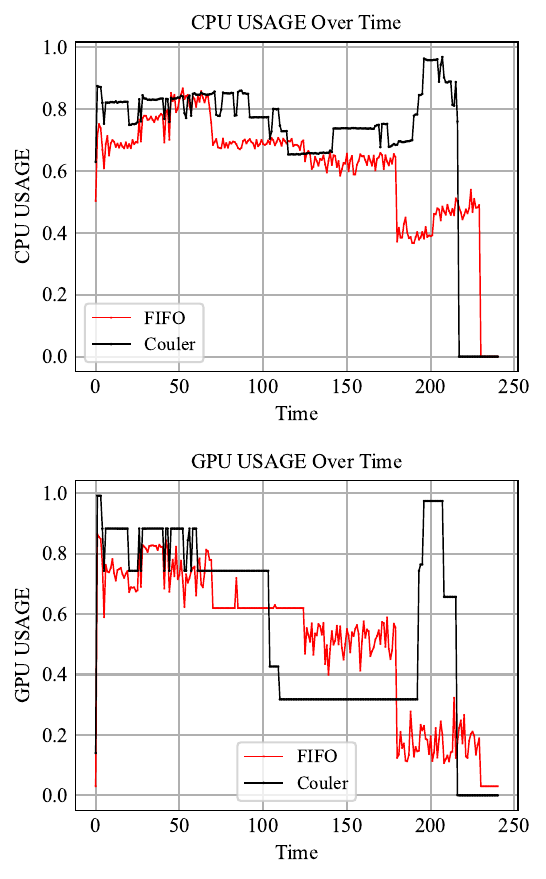}
        \caption{FIFO}
        \label{fig:Figure_NLP_FIFO}
    \end{subfigure}
    \hfill  
    \begin{subfigure}[b]{0.24\textwidth}  
        \includegraphics[width=\textwidth]{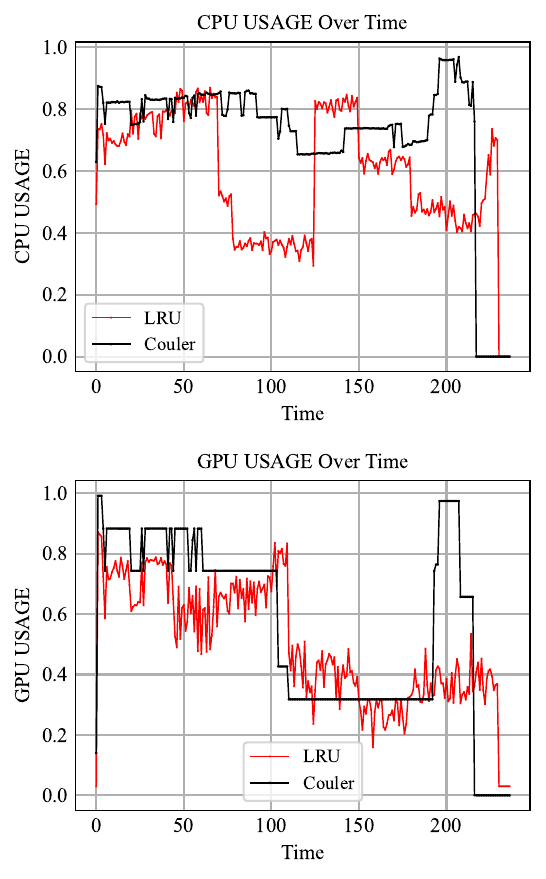}
        \caption{LRU}
        \label{fig:Figure_NLP_LRU}
    \end{subfigure}
    \caption{
Effect of \system on Resource Utilization and Workflow Execution Time in Language Model Fine-tuning Scenarios with FIFO and LRU Strategies}
    \label{fig:cache_perf2}
    \vspace{-1.0em}
\end{figure}

\begin{figure}[ht]
    \centering
    \begin{subfigure}[b]{0.24\textwidth}  
        \includegraphics[width=\textwidth]{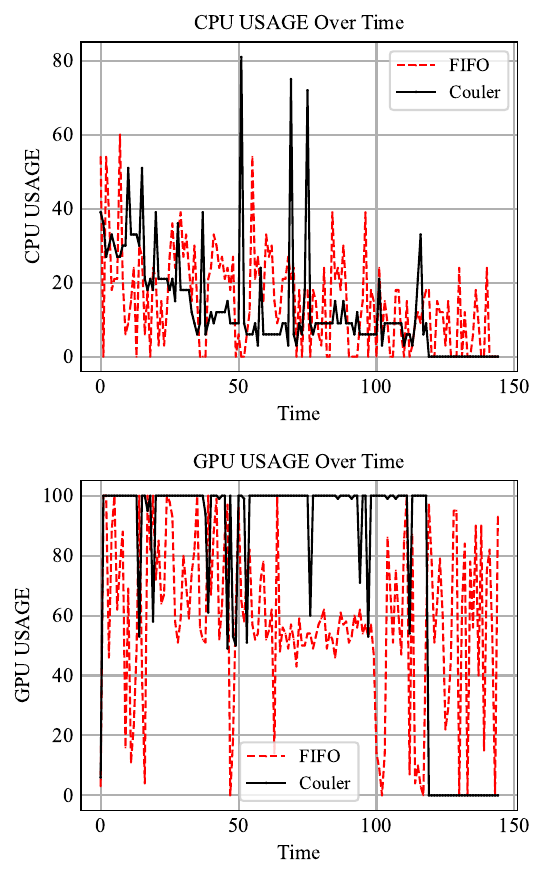}
        \caption{FIFO}
        \label{fig:Figure_multimodel_fifo}
    \end{subfigure}
    \hfill  
    \begin{subfigure}[b]{0.24\textwidth}  
        \includegraphics[width=\textwidth]{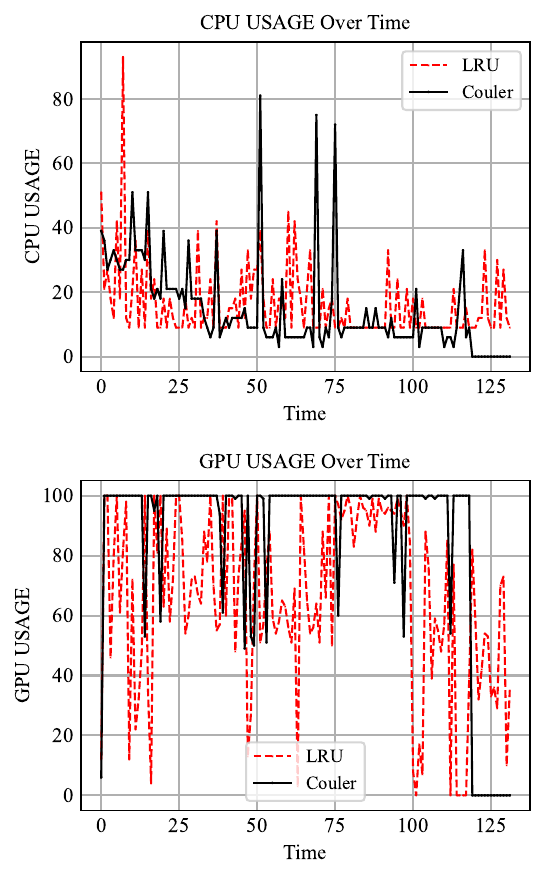}
        \caption{LRU}
        \label{fig:Figure_multimodel_lru}
    \end{subfigure}
    \caption{Effect of \system on Resource Utilization and Workflow Execution Time in Multimodal Training Scenarios with FIFO and LRU Strategies}
    \label{fig:cache_perf3}
    \vspace{-1.0em}
\end{figure}

\subsection{Performance Study with Cache Sizes}

Cache size is a significant factor affecting caching effectiveness, meaning that discussing the performance of the \system caching strategy under more limited cache sizes is valuable. This section details the changes in CPU/GPU utilization when the cache size is set to 10G and 20G, compared to a more ample 30G. It is observed that when the cache space is smaller, the weight of the Artifact caching cost increases, and simultaneously, some artifact units no longer meet the conditions for caching, leading to changes in caching situations, which in turn alters the workflow operation. When the cache size is more limited, the effectiveness of \system decreases, but overall, satisfactory results are achieved.

\begin{figure}[ht]
    \centering
    \begin{subfigure}[b]{0.24\textwidth}  
        \includegraphics[width=\textwidth]{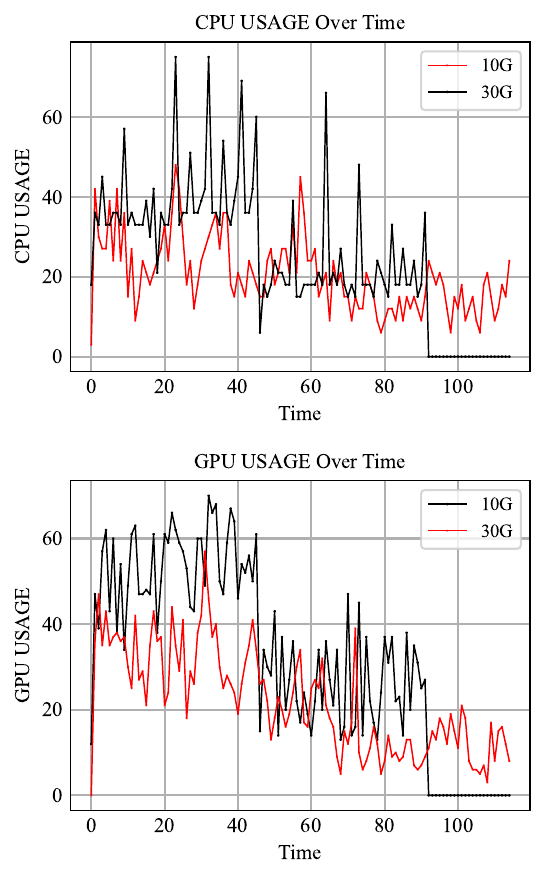}
        \caption{10G and 30G}
        \label{fig:Figure_CV_10G}
    \end{subfigure}
    \hfill  
    \begin{subfigure}[b]{0.24\textwidth}  
        \includegraphics[width=\textwidth]{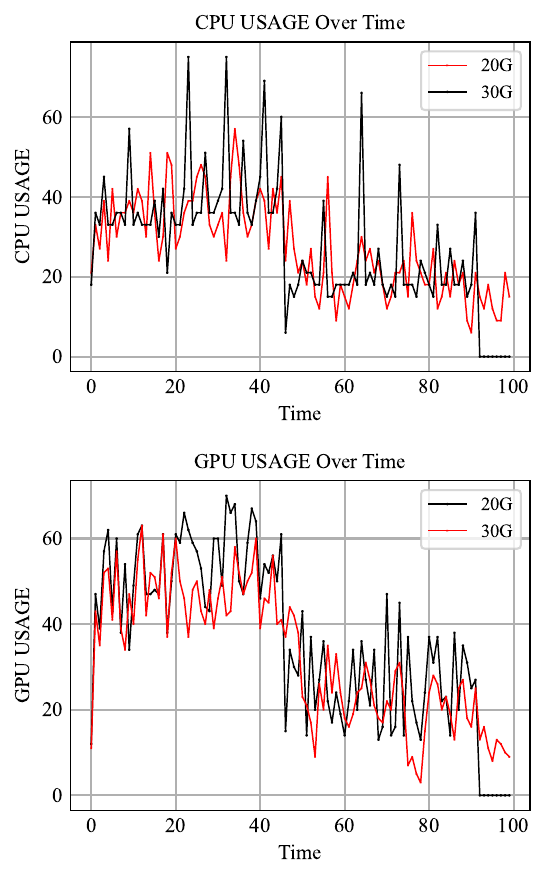}
        \caption{20G and 30G}
        \label{fig:Figure_CV_20G}
    \end{subfigure}
    \caption{Effect of \system on Resource Utilization and Workflow Execution Time with Different Cache Sizes in Image Segmentation Scenarios}
    \label{fig:cache_perf4}
    \vspace{-1.0em}
\end{figure}

\begin{figure}[ht]
    \centering
    \begin{subfigure}[b]{0.24\textwidth}  
        \includegraphics[width=\textwidth]{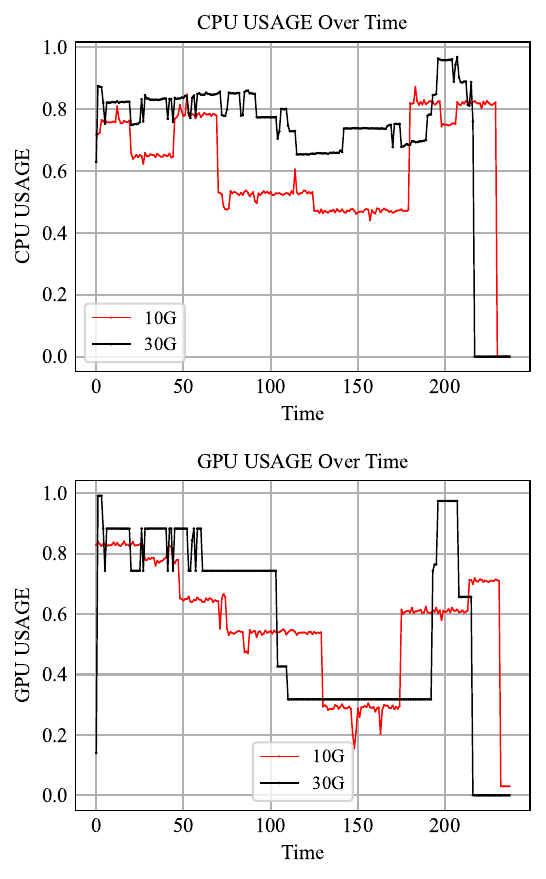}
        \caption{10G and 30G}
        \label{fig:Figure_NLP_10G}
    \end{subfigure}
    \hfill  
    \begin{subfigure}[b]{0.24\textwidth}  
        \includegraphics[width=\textwidth]{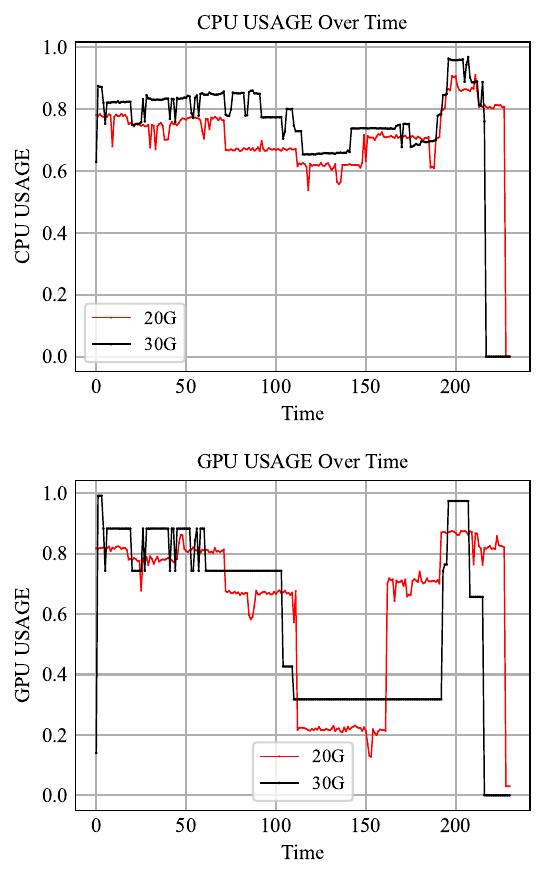}
        \caption{20G and 30G}
        \label{fig:Figure_NLP_20G}
    \end{subfigure}
    \caption{Effect of \system on Resource Utilization and Workflow Execution Time with Different Cache Sizes in Language Model Fine-tuning Scenarios}
    \label{fig:cache_perf5}
    \vspace{-1.0em}
\end{figure}

\begin{figure}[ht]
    \centering
    \begin{subfigure}[b]{0.24\textwidth}  
        \includegraphics[width=\textwidth]{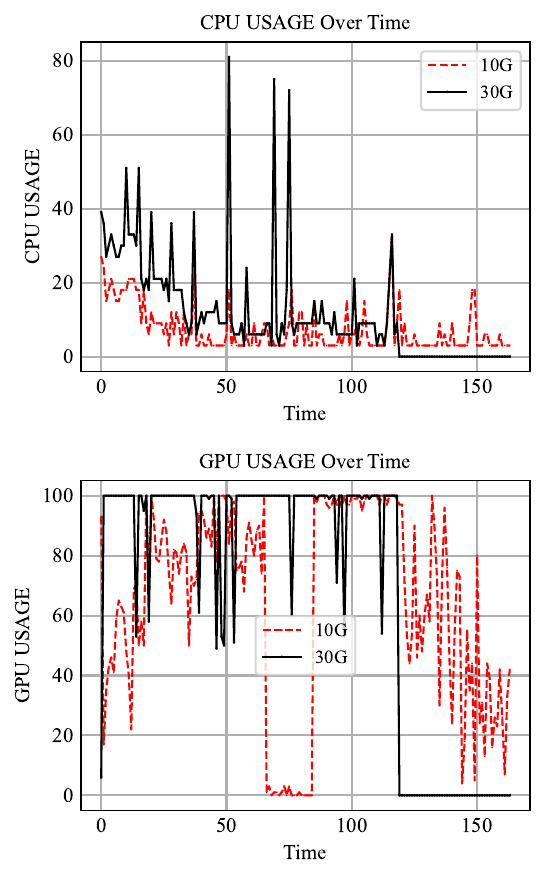}
        \caption{10G and 30G}
        \label{fig:Figure_multimodel_10G}
    \end{subfigure}
    \hfill  
    \begin{subfigure}[b]{0.24\textwidth}  
        \includegraphics[width=\textwidth]{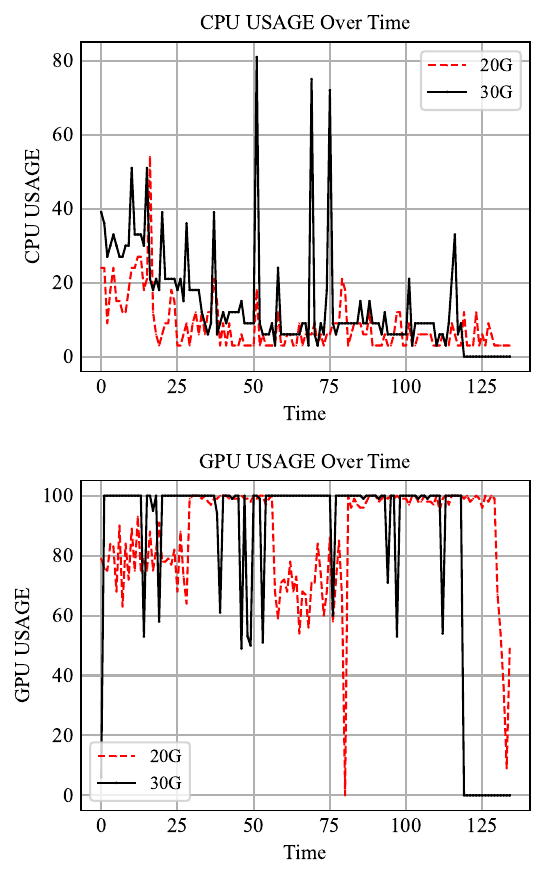}
        \caption{20G and 30G}
        \label{fig:Figure_multimodel_20G}
    \end{subfigure}
    \caption{Effect of \system on Resource Utilization and Workflow Execution Time with Different Cache Sizes in Multimodal Training Scenarios}
    \label{fig:cache_perf6}
    \vspace{-1.0em}
\end{figure}

\subsection{Performance study with Data caching}

\begin{figure}[ht]
    \centering
    \begin{subfigure}[b]{0.24\textwidth}  
        \includegraphics[width=\textwidth]{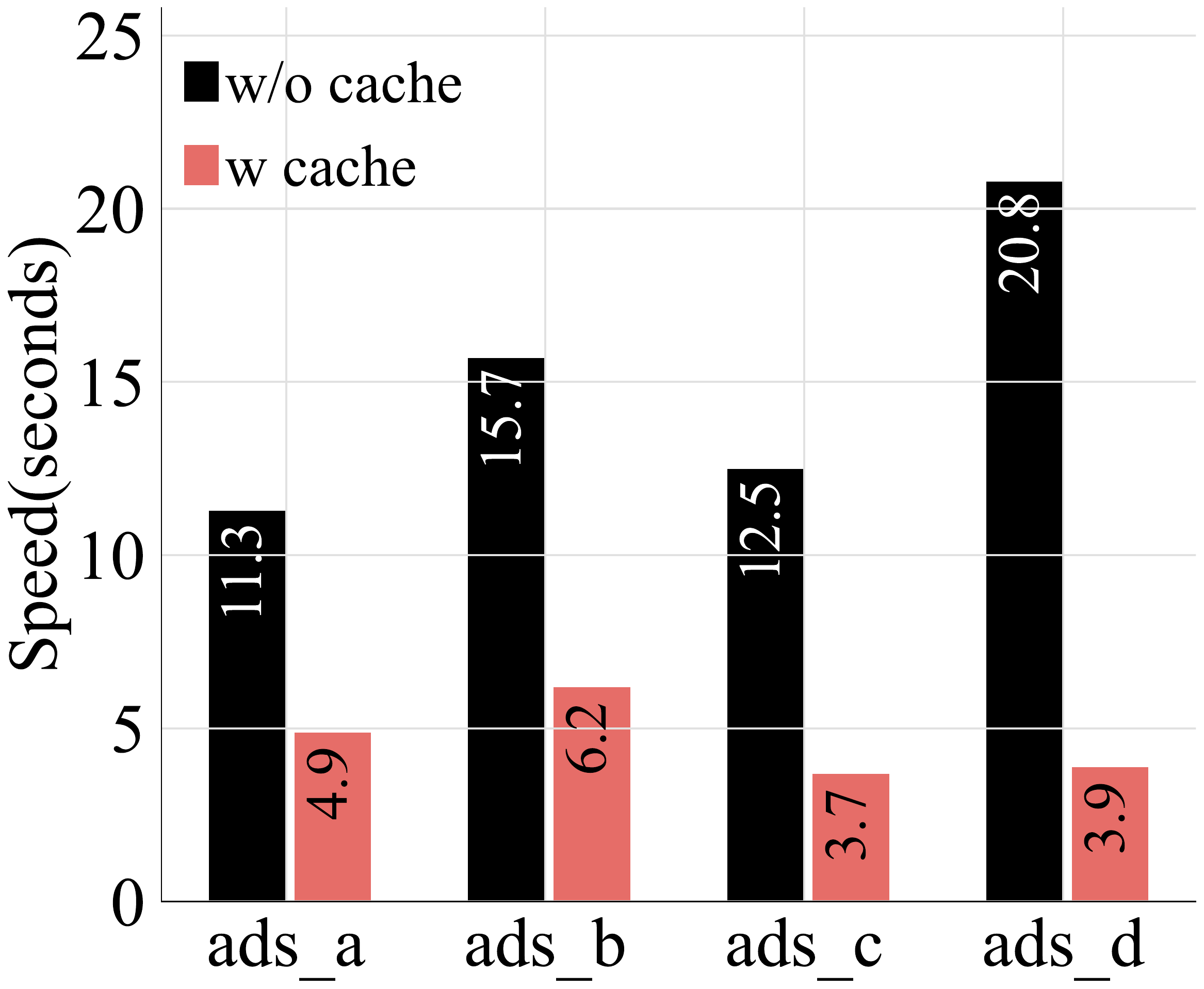}
        \caption{Table cached}
        \label{fig:table_cache}
    \end{subfigure}
    \hfill  
    \begin{subfigure}[b]{0.24\textwidth}  
        \includegraphics[width=\textwidth]{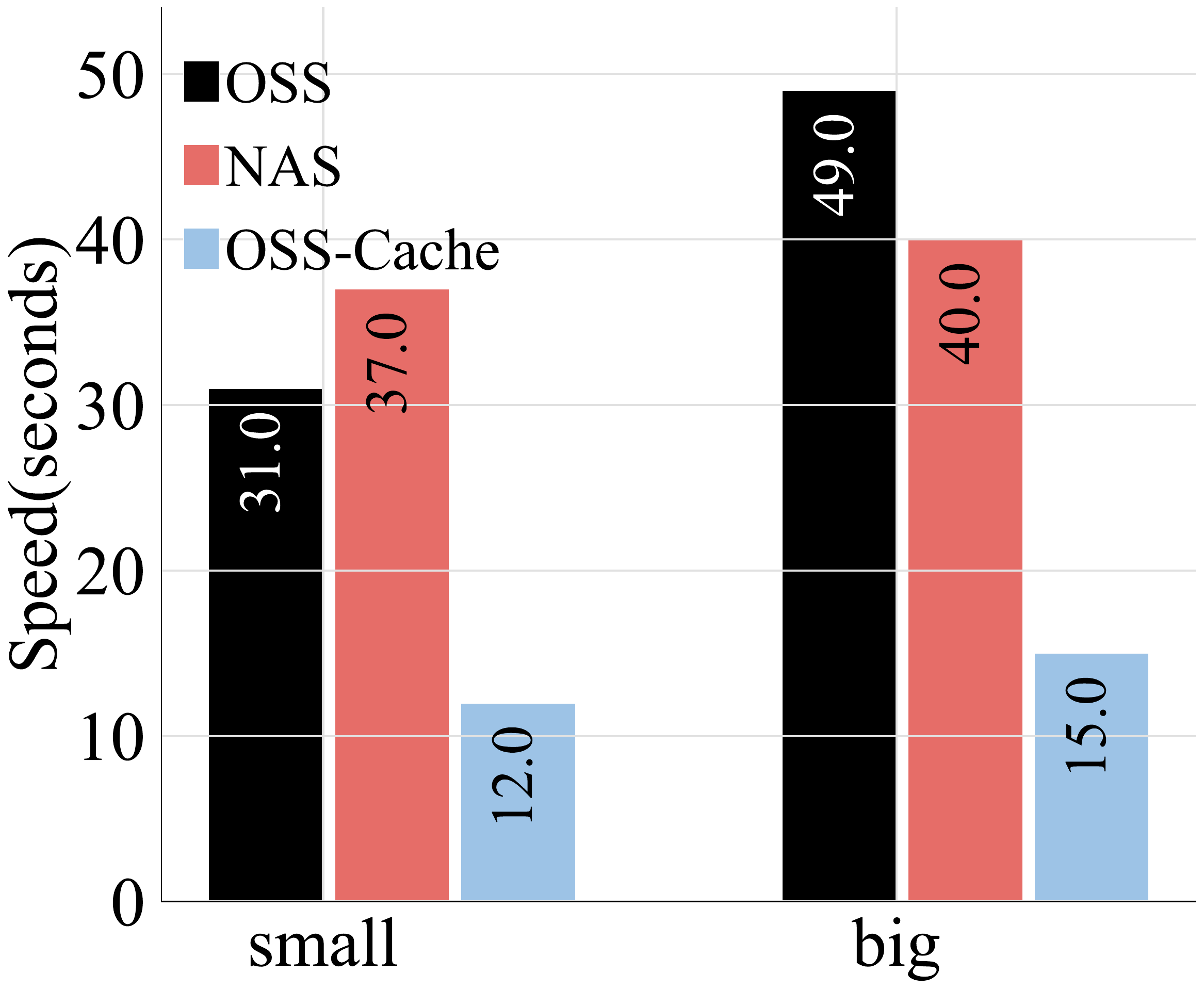}
        \caption{File cached}
        \label{fig:file_cache}
    \end{subfigure}
    \caption{Performance of data caching}
    \label{fig:cache_perf}
    \vspace{-1.0em}
\end{figure}

We study how the cache improves the data reading performance. 
At first, we use the two tables (e.g., ads-a and ads-b) from the ads recommendation application used in internal, the data is partitioned and stored in the Alibaba ODPS~\cite{ODPS} with an approximate size of bigger than 10GB per partition. We show how the cache improves the data loading performance, as well as how the caching to improves the deep learning model training over CPU and GPU configuration. The cluster is a hybrid model with offline and online server computation in the same computation node. The reading Pod is configured with an 8 core CPU and 8 GB memory to test the data reading throughput, then the deep learning model is configured with 10 parameter servers and 20 workers. From Figure~\ref{fig:table_cache}, we can observe the cache can improve the data loading performance twice, this confirms the local storage can reduce the cost of remote network data accessing.

Similarly, we also study the performance of the caching for small and big file reading. Initially, these files are stored in the remote file system (Alibaba OSS~\cite{OSS} and NAS~\cite{NAS}). For the small files application, the number of files is more than 10k and the total size of files is more than 10GB. The size of a big file is more than 1GB with .zip format, and the total number of files is more than 10. We test the caching performance based on the different number of jobs. From Figure~\ref{fig:file_cache}, we observe the local cache would improve the data reading speed more than 4 times comparing the data without cache.

\section{Production insights of \system}
Our work is fundamentally motivated by the challenges highlighted in earlier research~\cite{xin2021production} in Google GCP, specifically the observation of significant computational waste in machine learning (ML) workflows. This previous work provides a thorough analysis of 3000 ML production pipelines, revealing critical insights such as coarse-grained characteristics of these pipelines and the introduction of model graphlets. This analysis identified a crucial problem: a substantial amount of computation in these workflows does not lead to model deployment, thereby representing a significant inefficiency. This issue has emerged in our production environment as well, and we've noted similar observations about the workflow.

Building upon these findings, our contribution is directly aimed at addressing these inefficiencies. Our work revolves around designing a system for unified ML workflow optimization in the cloud. The contributions of our work are multifaceted: Simplicity and Extensibility, Automation, Efficiency and Real-World Impact and Adoption.

As for the insights from production in our experimental results, we believe the real-world adoption and application of our system within Ant Group and other companies serve as a testament to its practical value and effectiveness in production environments. The feedback and data obtained from these implementations have been crucial in refining our system to ensure it addresses the real-world challenges identified in previous studies.For example, the design of the workflow building block is as intuitive as procedural coding, inspired by real-world application workflow management. Similarly, the issue with large workflows is also driven by practical applications. Previously, we assumed that workflow sizes, and consequently the related YAML sizes, would not exceed 2MB. However, we've found that data scientists aim to construct very large workflows with hundreds of nodes and associated working procedures. This has directly influenced us to divide large workflows into smaller ones.

In conclusion, our work not only builds on the problems identified in earlier research but also provides tangible, applied solutions that have demonstrated real-world effectiveness. We hope this clarifies our contributions and addresses your concerns.
\end{appendices}

\end{document}